\documentclass[conference,letterpaper]{IEEEtran}

\usepackage{xcolor}
\usepackage[utf8]{inputenc} 
\usepackage[T1]{fontenc}
\usepackage{url}
\usepackage{ifthen}
\usepackage{cite}
\usepackage{comment}
\usepackage[cmex10]{amsmath} 
\usepackage{graphicx}

\usepackage{amsthm}
\usepackage{amsmath}
\usepackage{amsmath,amssymb,amsfonts}

\newtheorem{theorem}{Theorem}
\newtheorem{lemma}{Lemma}
\newtheorem{remark}{Remark}
\newtheorem{definition}{Definition}

\newcommand{\off}[1]{}

\allowdisplaybreaks
\usepackage{cuted}
\begin{document}
\title{Coding-Based Hybrid Post-Quantum Cryptosystem for Non-Uniform Information\vspace{-0.3cm}} 



\author{%
  \IEEEauthorblockN{Saar Tarnopolsky and
                    Alejandro Cohen}
  \IEEEauthorblockA{
                   Faculty of Electrical and Computer Engineering, Technion --- Institute of Technology, Haifa, Israel,\\Emails: saar@campus.technion.ac.il and alecohen@technion.ac.il\vspace{-0.45cm}}
}

\maketitle

\begin{abstract}
    We introduce for non-uniform messages a novel hybrid universal network coding cryptosystem (NU-HUNCC) in the finite blocklength regime that provides Post-Quantum (PQ) security at high communication rates. Recently, hybrid cryptosystems offered PQ security by premixing the data using secure coding schemes and encrypting only a small portion of it, assuming the data is uniformly distributed. \off{While by premixing the data using secure coding schemes, recently, hybrid cryptosystems\off{ against an all-observing eavesdropper} offered PQ security even if only a small portion of the data is encrypted, raw data messages were assumed uniformly distributed.} An assumption that is often challenging to enforce. Standard fixed-length lossless source coding and compression schemes guarantee a uniform output in \emph{normalized divergence}. Yet, this is not sufficient to guarantee security. We consider an efficient almost uniform compression scheme in \emph{non-normalized variational distance} for the proposed hybrid cryptosystem, that by utilizing a uniform sub-linear shared seed, guarantees PQ security.
    \off{We introduce a novel hybrid universal network coding cryptosystem for non-uniform messages (NU-HUNCC), providing a high-rate post-quantum (PQ) secured coding scheme in the final blocklength regime. Previous work introduced the HUNCC. By using an information-theoretically secured encoder and partial encryption, HUNCC guarantees PQ secure communication against an all-observing eavesdropper. The security of  HUNCC relies on uniformly distributed source messages, an assumption that is hard to enforce in practice.
    Although standard fixed-length lossless source coding and compression schemes guarantee a uniform output in normalized divergence, this is not enough to obtain security. We consider a special compression scheme, that by utilizing a uniform sub-linear shared seed, guarantees a uniform output in variational distance.}
    Specifically, for the proposed PQ cryptosystem, first, we provide an end-to-end coding scheme, NU-HUNCC, for non-uniform messages. Second, we show that NU-HUNCC is information-theoretic individually secured (IS) against an eavesdropper with access to any subset of the links. Third, we introduce a modified security definition, individually semantically secure under a chosen ciphertext attack (ISS-CCA1), and show that against an all-observing eavesdropper, NU-HUNCC satisfies its conditions. Finally, we provide an analysis that shows the high communication rate of NU-HUNCC and the negligibility of the shared seed size.
\end{abstract}

\section{Introduction}
We consider the problem of high data rate Post-Quantum (PQ) secure communication over a noiseless multipath network for non-uniform messages. In this setting, the transmitter, Alice, wants to send confidential non-uniform messages to the legitimate receiver, Bob, over noiseless communication links. We consider the two 
traditional eavesdroppers \cite{goldwasser2019probabilistic,forouzan2015cryptography,bloch2011physical}: 1) Information-Theory Eve (IT-Eve) which has access only to any subset of the communication links, but possess unlimited computational power \cite{bloch2011physical}, and 2) Cryptographic Eve (Crypto-Eve) which has access to all of the links, but is limited computationally \cite{goldwasser2019probabilistic}. We aim to ensure: 1) reliable communication between Alice and Bob, 2)  information-theoretic security against IT-Eve, and 3) PQ computational cryptographic security against Crypto-Eve \cite{bernstein2017post}.

Information-theoretic and computational cryptographic security are products of Shannon's seminal work on perfect secrecy \cite{Shannon1949} from 1949.
Shannon showed that perfect secrecy requires Alice and Bob to share a secret key with entropy higher or equal to the confidential message. Using such keys is costly and often non-practical. Information-theoretic and computational cryptographic security offer more practical solutions for secure communication at the expense of some relaxation on the security conditions. Traditional information-theoretic security relies on the probabilistic nature of communication channels, whereas computational cryptographic security assumes the eavesdropper has limited computational power. 

Wyner's work in 1975 on the wiretap channel \cite{WiretapWyner} introduced an information-theoretic secured coding scheme taking advantage of the eavesdropper's weaker observations of the transmissions over the channel. However, Wyner's approach required the use of a local source of randomness and a significant decrease in the communication rate \cite{bloch2011physical,liang2009physical,liang2009information,zhou2013physical}. In 1977, Carliel et al. \cite{carleial1977note} managed to increase the communication rate compared to Wyner's coding scheme, by replacing Wyner's source of local randomness with another uniform confidential message. This approach was named individual secrecy (IS). IS ensures IT-Eve is ignorant about any single message but may obtain some insignificant information about the mixture of all the messages. \off{The information leaked to IT-Eve in IS communication systems is usually measured by the non-normalized mutual information or the variational distance between each source message and IT-Eve's observations \cite{SecrecyChannelResolvability}.} Recently, IS was considered in various communication models such as the broadcast channel, terahertz wireless, etc \cite{JointIndividual2014,IndividualDegradedMU2015,SecrecyBroadcast,tan2019can,cohen2023absolute,yeh2023securing}, to efficiently increase data rates.

On the other hand, computational cryptographic security relies solely on the limited computational power of the eavesdropper. To securely transmit a message to Bob, Alice encrypts the message using a one-way function that is considered hard to invert. Thus, Crypto-Eve can't invert the function and decrypt the message in a reasonable time frame, but with the right shared key, inversion is possible for Bob \cite{forouzan2015cryptography}. Recent developments in quantum computing significantly increased Crypto-Eve's computational capabilities, rendering some of the most commonly used encryption schemes, unsecured \cite{QuantumRSA}. One of the public-key cryptosystems that to this day remain cryptographically secured against quantum computer-based algorithms is McEliecce \cite{mceliece1978public} and his probabilistic extensions, e.g.,  \cite{nojima2008semantic,dottling2012cca2,aguirre2019ind}. McEliecce is a coding-based cryptosystem first introduced in 1978. The security of McEliecce relies on the hardness of decoding a general linear code which is an NP-hard problem\off{decoding general linear codes, a problem considered NP-hard} \cite{berlekamp1973goppa,patterson1975algebraic}. However,\off{ The PQ cryptographic secure} McEliecce cryptosystem, inherently, suffers from low data rates of around $0.51$. Increasing the data rate of McElicce is possible, yet it is considered unsecured without a significant increase in its key size \cite{faugere2013distinguisher}.

In a work by A. Cohen et al. \cite{HUNCC}, a novel approach combining information theory and computational cryptographic security is introduced and named HUNCC. HUNCC combines an IS channel coding scheme \cite{SMSM}, that premixes the messages, with a PQ cryptographic secure scheme that encrypts a small amount of the mixed data. The result is a PQ secure communication scheme against both IT-Eve and Crypto-Eve attaining a data rate approaching one. In later works \cite{cohen2022partial,d2021post}, HUNCC was extended to support secure communication over noisy channels. However, the security of the IS coding scheme, and therefore also of HUNCC, relies on uniform source messages, a requirement often hard to enforce.

To guarantee security, the information leaked by HUNCC is measured by the \emph{non-normalized mutual information/variational distance} between Eve's observations and the source messages to demonstrate that this leakage meets the cryptographic criterion for ineligibility.  Although, non-uniform messages can be pre-compressed, standard lossless fixed-length source coding guarantees uniformity in \emph{normalized KL-divergence}, but not in \emph{non-normalized distances} \cite{han2005folklore}. \off{In 2013 Chou et al. \cite{chou2013data} showed that a lossless fixed-length source encoder requires a shared uniform seed of sub-linear size to guarantee its output is uniform in variational distance. In further works by Chou et al. \cite{chou2015polar}\cite{NegligbleCost} a polar codes-based source encoder for almost uniform output was suggested. By using a shared seed with negligible yet sub-optimal size, almost uniform messages are obtained. This source coding technique can be used to ensure strong secrecy over a wiretap channel by encoding the public message using an almost uniform source encoder and using it as a source of local randomness \cite{NegligbleCost}. In this setting, strong secrecy is obtained in the expanse of confidential data rate.} 

In this paper, we introduce a novel hybrid universal network coding cryptosystem for non-uniform messages (NU-HUNCC). First, we propose a source and IS channel coding scheme based on techniques presented in \cite{chou2013data,chou2015polar,NegligbleCost} and \cite{SMSM}, respectively. Against IT-Eve, PQ IS communication at a high data rate approaching capacity is obtained for non-uniform messages using a shared uniform seed of sub-linear size in an efficient pre-compression stage and then mixing the almost uniform outcome information. This result has an independent interest in information-theoretic security. Second, against Crypto-Eve, we present a new strong definition for PQ computational cryptography security, individual semantic security against chosen ciphertext attack (ISS-CCA1). We show that the controlled information leakage from the proposed source and IS channel coding scheme for non-uniform messages \emph{meets the cryptographic criterion for ineligibility}. Finally, we provide NU-HUNCC's data rate analysis against Crypto-Eve and show it can approach the capacity of the network under ISS-CCA1.

\section{System Model} \label{sec:sys}
We consider a communication system where Alice wishes to transmit $\ell$ non-uniform confidential message\footnote{In this paper, we assume the messages are independent to focus on the key methods and results. However, our proposed solution can be easily shown to be valid for dependent sources by using joint source coding schemes \cite{slepian1973noiseless}.} over $\ell$ noiseless links, $\mathcal{L}=\{1,...,\ell\}$, to Bob, in the presence of an eavesdropper, Eve. The messages are taken from a DMS $\left(\mathcal{V},p_V\right)$ s.t. $\mathcal{V} \in \{0,1\}$. We denote the source message matrix by $\underline{V}_{\mathcal{L}} \in \mathbb{F}_{2}^{\ell \times n}$ when $n$ is the size of each source message.

Bob's observations are denoted by $\underline{Y}_{\mathcal{L}}$. Those observations, provide Bob reliable decode $\underline{V}_{\mathcal{L}}$ with high probability. That is, $\mathbb{P}(\underline{\hat{V}}_{\mathcal{L}}(\underline{Y}_{\mathcal{L}}) \neq \underline{V}_{\mathcal{L}}) \leq \epsilon_e$, where $\underline{\hat{V}}_{\mathcal{L}}(\underline{Y}_{\mathcal{L}})$ is the estimation of $\underline{V}_{\mathcal{L}}$. We consider two types of Eve: 1) IT-Eve, which observes any subset $\mathcal{W} \subset \mathcal{L}$ of the links s.t. $|\mathcal{W}| \triangleq w < \ell$, but is computationally unbounded, and 2) Crypto-Eve which observes all the links, but is bounded computationally. We denote IT-Eve's observations by $\underline{Z}_{\mathcal{W}}$ and Crypto-Eve's observations by $\underline{Z}_{\mathcal{L}}$.

\off{
\begin{remark}
    To focus the paper on our main contributions we chose to work with the assumption that the messages are independent. However, our proposed solution can be easily shown to be valid for dependent sources by using joint source coding schemes \cite{slepian1973noiseless}. 
\end{remark}}

\subsection{Security against IT-Eve}
For any subset of $k_s < \ell - w$ source messages, we use the notion of $k_s$ individual security ($k_s$-IS) against IT-Eve. We measure the leakage of information to the eavesdropper using \emph{non-normalized variational distance}, denoted by $\mathbb{V}(\cdot,\cdot)$.
\begin{definition} \label{def:IS}
    ($k_s$ Individual Security) Let $\underline{V}_{\mathcal{L}} \in \mathbb{F}^{\ell \times n}_{q}$ 
    be a set of $\ell$ confidential source messages Alice intends to send, $\underline{Y}_{\mathcal{L}}$ be Bob's observations of the encoded messages, and $\underline{Z}_{\mathcal{W}}$ be IT-Eve's observations of the encoded messages. We say that the coding scheme is $k_s$-IS if $\forall \epsilon_s > 0$, $\forall \mathcal{W} \subset{\mathcal{L}}$ s.t. $|\mathcal{W}| = w < \ell$, and $\forall \underline{V}_{\mathcal{K}_s} \subset \underline{V}_{\mathcal{L}}$ s.t. $|\mathcal{K}_s|=k_s < \ell - w$, it holds that $\mathbb{V}(p_{\underline{Z}_{\mathcal{W}}|{\underline{V}_{\mathcal{K}_s}=\underline{v}_{\mathcal{K}_s}}},p_{\underline{Z}_{\mathcal{W}}}) \leq \epsilon_s$.
\end{definition}
Thus, IT-Eve that observes any subset of $w$ links in the network can't obtain any information about any set of $k_s$ individual messages, $\underline{V}_{\mathcal{K}_s}$. IT-Eve might be able to obtain some insignificant information about the mixture of all the messages. Yet, this information is controlled \off{and can't be used to learn anything about any set of $k_s$ individual messages} \cite{SMSM,SCMUniform,bhattad2005weakly}. 

\subsection{Security against Crypto-Eve}
Against Crypto-Eve we introduce a new notion of security, individual semantic security against a chosen ciphertext attack (ISS-CCA1). This notion of security is based on SS-CCA1 cryptographic security \cite{goldwasser2019probabilistic,bellare1998relations}, and usually requires the encryption scheme to be probabilistic\footnote{\label{comm:pub_key} To focus on our main contributions, we choose to work with public key encryption schemes, but any other PQ probabilistic encryption schemes would have achieved similar results.}.

\off{
\begin{remark}
     Probabilistic encryption schemes usually refer to public key encryption. However, there are symmetric key encryption algorithms that are also considered probabilistic. In this paper, to focus on our main contribution, we chose to work with public key encryption schemes, but any other probabilistic encryption schemes could have achieved similar results.
\end{remark}}
We start by defining public key cryptosystems.

\begin{definition}\label{def:Public-key}
    A public key cryptosystem consists of three algorithms: 1) Key generation algorithm $Gen(c)$ with an input $c$ which generates a public key, $p_c$, and a secret key, $s_c$. 2) An Encryption algorithm used by Alice which is taking a message $m$ and the public key, $p_c$ as an input, and outputs the ciphertext $c$. We denote the encryption algorithm as $\kappa = Crypt(m,p_c)$. 3) A polynomial time decryption algorithm taking the ciphertext, $\kappa$, and the secret key, $s_c$, and output the original message $m = Crypt^{-1}(Crypt(m,p_c),s_c)$.
\end{definition}

The notion of semantic security (SS) in computational cryptographic security \off{was first introduced in \cite{goldwasser2019probabilistic} and} is \off{considered to be} the equivalent of Shannon's perfect secrecy against a computationally bounded eavesdropper\cite{goldwasser2019probabilistic}. The definition of SS-CCA1 can be found in \cite{bellare1998relations,dowsley2009cca2,aguirre2019ind}. We introduce a new notion of semantic security based on the definition of SS-CCA1, Definition~\ref{def:IS}, and Definition~\ref{def:Public-key}:  

\begin{definition}\label{def:individuall-SS-CCA1}
    (ISS-CCA1). Individual semantic security under a chosen ciphertext attack is defined by the following game between an adversary and a challenger:
        1) The challenger generates a key pair using a security parameter $c$: $Gen(c) = (p_c,s_c)$ and shares $p_c$ with the adversary.
        2) The adversary outputs a pair $(\mathcal{M},S)$, where $\mathcal{M} \in \mathbb{F}_{q}^{\ell}$ is the messages space, and $S$ is a state.
        3)  The adversary sends a polynomial number of ciphertexts to the challenger that decrypts them and returns them to the adversary.
        4) The adversary chooses an index $i^* \in \{1,...,\ell\}$ and tells it to the challenger.
        5) Let $\mathcal{F} = \{f : \mathcal{M} \rightarrow \Sigma\}$ be the set of functions on the message space. For any value $\sigma \in \Sigma$ we denote by $f^{-1}(\sigma)$ the inverse image of $\sigma$ on the message space, $\{m \in \mathcal{M}|f(m)=\sigma \}$. We denote the probability for the most probable value for $f(m)$ as $p_{\sigma-max} = \max\limits_{\sigma \in \Sigma}\{ \sum_{m\in f^{-1}(\sigma)} p(m) \}$.
        6) For every $i \in \{1,...,\ell\}$, we consider $m_i$ as an individual message. The challenger samples each $m_i$ from its probability distribution creating the message $m$. 
        7) The challenger encrypts all the messages $\kappa = Crypt(m,p_c)$ and sends it to the adversary.
        8) The adversary tries to find a pair $(f,\sigma)$ such that $\sigma = f(m_{i^*})$. 
    If the adversary finds a pair $(f,\sigma)$ s.t. $\sigma=f(m_{i^{*}})$ he wins. The cryptosystem is considered ISS-CCA1 if for any pair $(\mathcal{M},S)$ and for any function $f \in \mathcal{F}$, the advantage the adversary has over guessing according to the message space probability distribution is negligible. We denote a function $\epsilon(c)$ s.t. for every $d > 0 $, there exists an integer $c_d$ such that $\forall c > c_d$, the bound $\epsilon(c) < \frac{1}{c^d}$ holds. The function $\epsilon(c)$ is called a negligible function. The advantage of the adversary is measured by the adversary's probability to win: $p_{\sigma-max} + \epsilon(c)$. When $\epsilon(c)$ is a negligible function, the adversary's advantage is negligible as well.
\end{definition}
This definition of ISS-CCA1 guarantees that Crypto-Eve's advantage over guessing $m_i$ according to its distribution is negligible for all $1 \leq i \leq \ell$. 

\section{Non-Uniform Hybrid Universal Network Coding Cryptosystem (NU-HUNCC)}\label{sec:main_results}
\begin{figure*}[htbp]
  \centering
  \includegraphics[width=0.86\textwidth]{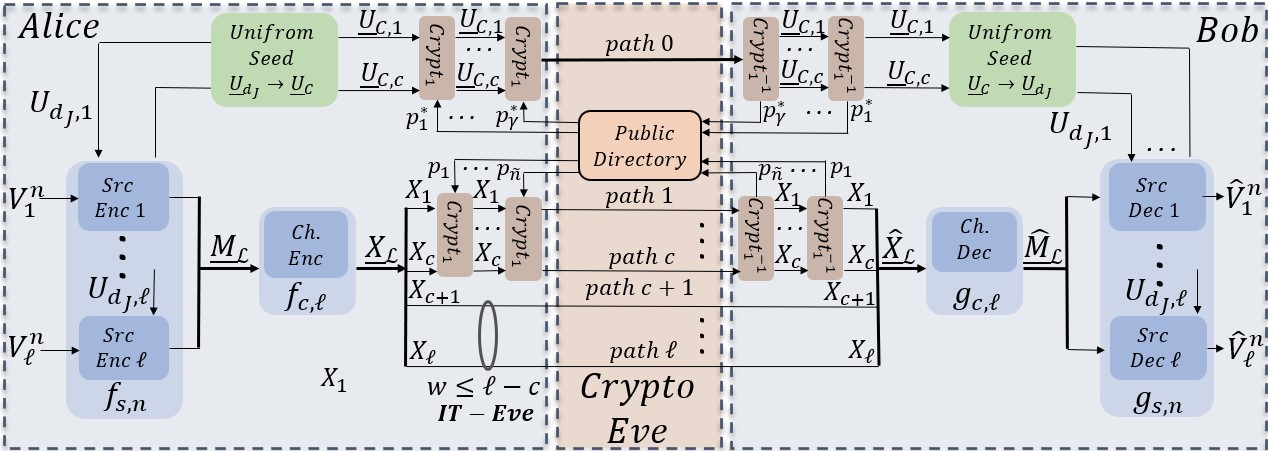}
  \vspace{-0.2cm}
  \caption{NU-HUNCC cryptosystem with $\ell$ noiseless communication links and two types of Eve's: IT-Eve with access to $w < \ell$ links, and Crypto-Eve with access to all the links. The lossless almost uniform compression is done by the polar codes-based encoder from \cite{NegligbleCost}.$c$ of the links are encrypted by a PQ public-key SS-CCA1 cryptosystem. The mixing of the messages is done by the individual secure random network coding scheme from \cite{SMSM,cohen2022partial}. The uniform seed is encrypted as well and shared by a separate link. In practice, the encrypted seed is concatenated to the $c$ encrypted messages.}
  \label{fig:NU-HUNCC}
  \vspace{-0.6cm}
\end{figure*}

In this section, we present our proposed NU-HUNCC scheme against IT-Eve and Crypto-Eve as illustrated in Fig.~\ref{fig:NU-HUNCC}. We present the main results of our work and provide a brief description of the PQ secure code construction.

NU-HUNCC provides a way to efficiently transmit non-uniform messages between Alice and Bob, by encrypting only a subset of the links and maintaining PQ security at high data rate communication. We present three main novelties: 1) An efficient end-to-end communication scheme for non-uniform messages, 2) The proposed scheme is $k_s$-IS against IT-Eve, and 3) ISS-CCA1 PQ secure against Crypto-Eve.

NU-HUNCC cryptosystem consists of three main parts: 1) A lossless almost uniform source coding scheme using a uniform seed shared efficiently with the source decoder, 2) A joint IS message channel coding scheme to pre-mix the almost uniform messages, and 3) An SS-CCA1 encryption scheme used to encrypt only $1 \leq c < \ell$ links against Crypto-Eve.\footnote{The uniform seed can be shared between Alice and Bob in the expanse of some rate. Against Crypto-Eve, the seed as we demonstrate in this section is shared over one of the encrypted links. Against IT-Eve, the seed can be shared using wiretap coding techniques \cite[Chapter 4]{bloch2011physical}.}

\off{
\begin{remark} \label{rm:seed_sharing}
    The uniform seed can be shared between Alice and Bob in the expanse of some rate. Against Crypto-Eve, the seed as we demonstrate in this section is shared over one of the encrypted links. Against IT-Eve, the seed can be shared using wiretap coding techniques \cite[Chapter 4]{bloch2011physical}.
\end{remark}}

Now we present our proposed NU-HUNCC scheme. Due to space limitations, we give here a brief description of the code construction. The full descriptions of the source and IS channel schemes are given in Appendix~\ref{appendix:src_code} and~\ref{appendix:msg_encoder}, respectively.

\underline{\textit{Source coding scheme}}: For the almost uniform source coding a specialized version of a polar-codes-based source encoder is used \cite[proposition 4]{NegligbleCost}. Let $m \in \mathbb{N}$, and the source messages blocklength be $n = 2^m$. First, each row of the source message matrix, $\underline{V}_{\mathcal{L}}$, is separately encoded by the polarization transformation \cite{ArikanBase2009}. Let $\delta_n \triangleq 2^{-n^\beta}$ for $\beta \in [0,\frac{1}{2})$. We denote a polarized message by $A \in \mathbb{F}_2^n$, and by using $\delta_n$ the set of $n$ polarized bits are divided into three groups: 1) The bits with entropy almost 1 - $\mathcal{H}_V \triangleq \{ j \in [1,n]: H(A^{(j)}|A^{j-1}) > 1 - \delta_n \}$, 2) The bits with entropy almost 0 - $\mathcal{U}_V \triangleq \{ j \in [1,n]: H(A^{(j)}|A^{j-1}) < \delta_n \}$, and 3) The complementary group of bits with entropy in-between - $\mathcal{J}_V = (\mathcal{U}_V \cup \mathcal{H}_V)^{C}$. The compressed message is formed from $\mathcal{H}_V \cup \mathcal{J}_V$, the bits with non-negligible entropy, i.e., groups 1 and 3. For the compressed message to be almost uniform, a uniform seed of size $|\mathcal{J}_{V}| \triangleq d_{J}$ must be used on each message separately. The uniform seed one-time pads the bits from group $\mathcal{J}_V$. We denote the random seed matrix for all the messages by $\underline{U}_{d_{J}} \in \mathbb{F}_{2}^{\ell \times d_{J}}$. The size of the almost uniform message is $\tilde{n} = |\mathcal{H}_{V}| + d_{J}$ and the encoder for the $i$-th row the source message matrix denoted by $f_{s,n}$, operates as follows 
\begin{gather} \label{eq:src_code}
    f_{s,n} : (\underline{V}_{\mathcal{L},i},\underline{U}_{d_{J},i}) \in \mathbb{F}_{2}^{n} \times \mathbb{F}_{2}^{d_{J}} \rightarrow \underline{M}_{\mathcal{L},i} \in \mathbb{F}_{2}^{\tilde{n}}
\end{gather}
\begin{remark} \label{rm:polar}
    The size of the seed can be bounded by $n^{0.7214} \leq d_{J} \leq n^{0.7331}$. A more detailed analysis \off{of the seed's size}is given in Appendix~\ref{appendix:seed_length}. \off{In \cite{ArikanBase2009} it has been shown that $\lim \limits_{n \rightarrow \infty} \frac{|\mathcal{H}_V|}{n} = H(V)$. Thus, By choosing $n$ large enough, the compression is close to its optimal rate while the seed size is negligible compared to the compressed message size.}
\end{remark}

\underline{\textit{IS channel coding scheme}}: \off{For the channel coding,} We use a random coding-based joint encoder as given in \cite[Section IV]{SMSM}. The channel code pre-mixes the messages by encoding the message matrix column by column.\off{The codebook is generated from an i.i.d. source $X ~ Bernouli(\frac{1}{2})$ s.t. the probability distribution of each codeword, $x^l$, is $P(x^{\ell}) = \prod_{i=1}^{\ell}P(x_i)$.} The random codebook is organized in a binning structure of $2^{k_s}$ bins s.t. each bin has $2^{k_w}$ codewords, where $k_w \triangleq \ell - k_s$. For each column in the message matrix, $k_s$ bits are used to choose a bin, and the other $k_w$ bits are used to choose a codeword inside the bin. Thus, we receive the codeword matrix $\underline{X}_{\mathcal{L}}$. The encoder of a column is denoted by $f_{c,\ell}$. The encoding of the $j$-th column operates as follows
\begin{gather} \label{eq:msg_code}
    f_{c,\ell} : \underline{M}_{\mathcal{L}}^{(j)} \in \mathbb{F}_{2}^{\ell} \rightarrow \underline{X}_{\mathcal{L}}^{(j)} \in \mathbb{F}_{2}^{\ell}
\end{gather}

Using the source and IS channel coding scheme against IT-Eve, we provide the following achievability theorem.

\begin{theorem} \label{Direct}
   Assume a noiseless multipath communication $(\ell,w)$. NU-HUNCC reliably delivers with high probability $\ell$ non-uniform messages from a DMS $(\mathcal{V},p_V)$ to the legitimate receiver, such that $\mathbb{P}(\underline{\hat{V}}_{\mathcal{L}}(\underline{Y}_{\mathcal{L}}) \neq \underline{V}_{\mathcal{L}}) \leq \epsilon_e$, while keeping IT-Eve ignorant of any set of $k_s \leq \ell - w - \ell\epsilon$ messages individually, such that $\mathbb{V}(p_{\underline{Z}_{\mathcal{W}}|{\underline{V}_{\mathcal{K}_s}=\underline{v}_{\mathcal{K}_s}}},p_{\underline{Z}_{\mathcal{W}}}) \leq \epsilon_s$, whenever $d_{J} \geq |\mathcal{J}_V|$, $\ell\epsilon = o(\ell)$, and $\ell$ is lower bounded by $\omega(\tilde{n}^{\frac{2}{t}})$ for some $t \geq 1$ and upper bounded by $o(2^{n^\beta}/\tilde{n})$. 
   \off{ The IS communication rate obtained is $R = \frac{1}{\frac{|\mathcal{H}_V|}{n} + \frac{2d_{J,n}}{n}}$.}
\end{theorem}
\off{\textit{Proof(Rate)}: Alice wants to send $\ell n$ symbols to Bob. The message to Bob contains the following information: 1) $\ell$ premixed messages of total size $\tilde{n} \cdot \ell$, 2) $\ell$ uniform seed realizations of size $d_{J,n}$ each. Dividing the numerator and denominator by $n$, gives the required result. \qed}

The achievability and leakage proofs against IT-Eve are given in Section~\ref{sec:IT-Eve} and Appendix~\ref{appendix:IT-secrecy}, respectively. We do note, that as demonstrated in these proofs, the number of messages $\ell$ in Theorem~\ref{Direct} is lower and upper bounded as a function of $n$, due to the reliability and security constraints, respectively. I.e., for $\mathbb{P}(\underline{\hat{V}}_{\mathcal{L}}(\underline{Y}_{\mathcal{L}}) \neq \underline{V}_{\mathcal{L}}) \leq  \tilde{n}O(2^{-\ell}) + \sqrt{2\tilde{n}\ell 2^{-n^{\beta}}} + \ell 2^{-n^{\beta}}$ and $\mathbb{V}(p_{\underline{Z}_{\mathcal{W}}|{\underline{V}_{\mathcal{K}_s}=\underline{v}_{\mathcal{K}_s}}},p_{\underline{Z}_{\mathcal{W}}}) \leq \tilde{n}\ell^{-\frac{t}{2}} + 2\sqrt{2\tilde{n}\ell 2^{-n^{\beta}}}$.
\vspace{0.05cm}

\off{\begin{remark}
   In \cite{NegligbleCost} it was shown that the polar codes-based source encoder could be used to achieve strong secrecy over a wiretap channel with non-uniform public and confidential messages. Our setting efficiently delivers $l$ confidential messages, keeping them $k_s$-IS by using a joint channel coding scheme.
    \end{remark}}

Now, we consider NU-HUNCC scheme against Crypto-Eve, which has access to all the noiseless links. Thus, we use an encryption scheme in addition to the source and IS scheme. We show that ISS-CCA1 (see Definition~\ref{def:individuall-SS-CCA1}) security against Crypto-Eve can be obtained by encrypting only a portion of the links, $1 \leq c < \ell$, and the uniform seed with an SS-CCA1 cryptosystem. \off{In addition, we require the the uniform seed used by the source encoder/decoder pair, to be encrypted as well and concatenated to one of the encrypted links.}In this paper, we consider the cryptosystem to be a public key$^{\ref{comm:pub_key}}$ cryptosystem (see Definition~\ref{def:Public-key}) operating on $c$ bits at a time and denote it $Crypt_1$. $Crypt_1$ maps each $c$ bits into $c+r$ bits s.t. the additional bits are the added randomness of size $r$ required for $Crypt_1$ to be SS-CCA1 \cite{goldwasser2019probabilistic}. The uniform seed matrix is rearranged and denoted by $\underline{U}_{C} \in \mathbb{F}_{2}^{c \times \gamma}$ s.t. each of the $\gamma \triangleq \frac{\ell d_J}{c}$ columns has $c$ symbols. $Crypt_1$ is employed on: 1) $c$ bits of each $j$-th column, $j\in\{1,\ldots,\tilde{n}\}$, of the encoded matrix $\underline{X}_{\mathcal{L}}$ which are denoted by $X^{(j)}$, and 2) each column $k\in\{1,\ldots,\gamma\}$ of the seed matrix, $\underline{U}_{C}^{(k)}$. Thus, for the $j$-th column of $\underline{X}_{\mathcal{L}}$ and $k$-th column of $\underline{U}_{C}$, we have\off{  operates on $c$ bits at a time: 1)  of each column of the encoded messages $\underline{X}_{\mathcal{L}}$ and the uniform seed matrix $\underline{U}_{d_{J}}$. $\underline{U}_{d_{J}}$ is and encrypted column by column For every $c$ bits being encrypted, $Crypt_1$ outputs $r$ additional bits. The additional $r$ bits at the output of the encryption scheme are due to the added randomness required from $Crypt_1$ for it to be SS-CCA1 \cite{goldwasser2019probabilistic} and are concatenated to the encrypted links along with the encrypted seed. Each $c$ symbols of the $j$-th column in the encoded message matrix, $\underline{X}_{\mathcal{L}}$ denoted by $X_j$, and each $c$ symbols from seed matrix, $\underline{U}_{d_{J}}$ denoted by $U_c$, are encrypted separately, s.t.}
\begin{multline} \label{eq:encryption}
    Crypt_1 : X^{(j)} \in \mathbb{F}^c_{2} \rightarrow Y^{(j)} \in \mathbb{F}^{c+r}_{2},\\ 
    \text{and } \underline{U}^{(k)}_{C} \in \mathbb{F}^c_{2} \rightarrow Y^{(\tilde{n}+k)} \in \mathbb{F}^{c+r}_{2},
\end{multline}
where the additional $r$ bits of $Y^{(j)}$ and the entire $Y^{(\tilde{n}+k)}$ are concatenated to the $c$ encrypted links between Alice and Bob.
\off{
Each $c$ bits from the uniform seed matrix, $\underline{U}_{d_{J}}$ are also encrypted separately s.t.
\begin{gather} \label{eq:encryption_seed}
    Crypt_1 : U_i \in \mathbb{F}^c_{2} \rightarrow U^{*}_i \in \mathbb{F}^{c+r}_{2}.
\end{gather}}
We denote the encryption of the $j$-th column of the message matrix $\underline{M}_{\mathcal{L}}$ as $Crypt_2$. s.t.
\begin{gather} \label{eq:encryption_crypt2}
    Crypt_2 : \underline{M}_{\mathcal{L}}^{(j)} \in \mathbb{F}^{\ell}_{2} \rightarrow \underline{Y}_{\mathcal{L}}^{(j)} \in \mathbb{F}^{\ell+r}_{2}.
\end{gather}
$Crypt_2$ combines the source and IS channel coding scheme with the PQ SS-CCA1 encryption scheme. We denote the encrypted links indexes as $\mathcal{C}$ and non-encrypted links indexes as $\mathcal{C}^{C} = \mathcal{L} \setminus \mathcal{C}$. Thus, for the $j$-th column Bob's observations are $ \underline{Y}_{\mathcal{L}}^{(j)} = [Crypt_1(f_{c,\ell}(\underline{M}_{\mathcal{L}}^{(j)})[\mathcal{C}],p_c),f_{c,l}(\underline{M}_{\mathcal{L}}^{(j)})[\mathcal{C}^{C}]]$, where $p_c$ is the public key for $Crypt_1$ as depicted in Definition~\ref{def:Public-key}.

We now give the theorem claiming our end-to-end coding and encryption scheme is secured against Crypto-Eve.
\begin{theorem} \label{Individual-SS-CCA1}
    Assume the setting in Theorem~\ref{Direct}. Let $Crypt_1$ be an SS-CCA1 secured cryptosystem as in (3)\off{ \eqref{eq:encryption}}. Then, NU-HUNCC is ISS-CCA1 secure.
\end{theorem}
\vspace{-0.15cm}
The security proof of the theorem against Crypto-Eve is given in Section~\ref{sec:Crypto-Eve} and Appendix~\ref{appendix:individual-ss-cca1}, respectively.

\underline{\textit{Decoding At Bob}}:
First, Bob decrypts each $c$ encrypted symbols separately (including the shared seed) using the secret key $s_c$\footnote{Against IT-Eve the decryption stage is not required.}. After obtaining the decrypted symbols, Bob directly decodes each column of the encoded message matrix by looking for: 1) The bin in which the codeword resides, and 2) The offset of the codeword inside its corresponding bin. We note that this can be done directly since the links are noiseless. Bob's channel decoding error probability is negligible as given in \cite[Section \uppercase\expandafter{\romannumeral4}]{SMSM}. Then, for each row separately, Bob one-time pads the bits from group $\mathcal{J}_V$ using the decrypted seed, and employs successive cancellation decoder as given in \cite{ArikanBase2009,cronie2010lossless} to reliably decode each source message. Considering the negligible error probability of each code separately it can be shown by applying the coupling lemma \cite[lemma 3.6]{aldous1983random} \off{and union bound} that the suggested coding scheme is reliable\off{By applying the coupling lemma \cite[lemma 3.6]{aldous1983random}, and the union bound, and considering the negligible error probability of each of the codes separately it can be shown the suggested coding scheme is reliable}. Due to space limitation, we refer to Appendix~\ref{appendix:reliability} for the full proof.

The data rate of NU-HUNCC with $1 \leq c < \ell$ encrypted links against Crypto-Eve is given by the following theorem.

\begin{theorem} \label{Crypto-Eve-Rate}
    Consider the setting of the NU-HUNCC with an encoding scheme as in Theorem~\ref{Direct}, the information rate of NU-HUNCC is
    \begin{gather*} \label{eq:Crypto-Eve-Rate}
        R = \frac{1}{\frac{|\mathcal{H}_V|}{n}(1 +\frac{r}{\ell})+ \frac{d_{J}}{n}(2+\frac{r}{\ell} + \frac{r}{c})}.
    \end{gather*}
\end{theorem}

\textit{Proof}: Alice wants to send $\ell n$ symbols to Bob. The messages to Bob contain the following information: 1) $\ell-c$ not encrypted messages of total size $\tilde{n}(\ell-c)$, 2) $c$ encrypted messages of total size $\tilde{n}(c+r)$, and 3) encrypted seed of size $\ell d_{J} \cdot \frac{c+r}{c}$. Dividing the numerator and denominator by $\ell n$, we obtain the result in Theorem~\ref{Crypto-Eve-Rate} on NU-HUNCC data rate. \qed
\off{
\begin{remark}
    Using the polarization properties (Remark \ref{rm:polar}), it can be shown that for $n$ large enough: $R = \frac{ell}{H(V) + \frac{r}{\ell}}$, where $r$ is the size of the randomness required to encrypt $c$ bits. 
\end{remark}}

\section{Secure Individual Random code against IT-Eve \\ (Security  Proof Sketch of Theorem~\ref{Direct})}\label{sec:IT-Eve}
In this section, we provide a security proof sketch for Theorem~\ref{Direct} against IT-Eve. Due to space limitations, we refer to Appendix~\ref{appendix:IT-secrecy} for the full security proof. 
We consider noiseless multipath communication as described in Section~\ref{sec:sys}. To transmit $\ell$ confidential messages, all from the source $(\mathcal{V},p_V)$, Alice employs the polar codes-based source encoder and channel encoder from Section~\ref{sec:main_results}. Here, we show that IT-Eve, observing any subset of $w$ links, $\underline{Z}_{\mathcal{W}}$, can't obtain any significant information about any set of $k_s$ messages individually. We denote a set of $k_s$ messages by $\mathcal{K}_s$, s.t. $|\mathcal{K}_s| \triangleq k_s \leq \ell - w - \ell\epsilon$. The uniformly distributed message matrix is denoted by $\underline{\tilde{M}}_{\mathcal{L}}$, and its distribution is denoted by $p_{U_{\underline{M}_{\mathcal{L}}}}$. The distribution of $\underline{Z}_{\mathcal{W}}$ induced from uniformly distributed messages is denoted by $\tilde{p}_{\underline{Z}_{\mathcal{W}}}$. Thus, using similar techniques given in \cite{NegligbleCost} for strong security in broadcast wiretap channels\footnote{In \cite{NegligbleCost}, it was shown that a polar codes-based source encoder could be used to achieve strong secrecy over a broadcast channel with non-uniform public and confidential messages. Our setting efficiently delivers $\ell$ confidential messages, keeping them $k_s$-IS by using a joint channel coding scheme.}, we have
\vspace{-0.15cm}
\begin{align}&\mathbb{V}\left(p_{\underline{Z}_{\mathcal{W}}|\underline{M}_{\mathcal{K}_s}=\underline{m}_{k_s}},p_{\underline{Z}_{\mathcal{W}}}\right) \leq  \mathbb{V}\left(\tilde{p}_{\underline{Z}_{\mathcal{W}}|\underline{M}_{\mathcal{K}_s}=\underline{m}_{k_s}},\tilde{p}_{\underline{Z}_{\mathcal{W}}}\right) \nonumber \\ 
    & + \mathbb{V}\left(p_{\underline{Z}_{\mathcal{W}}|\underline{M}_{\mathcal{K}_s}=\underline{m}_{k_s}},\tilde{p}_{\underline{Z}_{\mathcal{W}}|\underline{M}_{\mathcal{K}_s}=\underline{m}_{k_s}}\right) + \mathbb{V}\left(\tilde{p}_{\underline{Z}_{\mathcal{W}}},p_{\underline{Z}_{\mathcal{W}}}\right), \label{eq:k_s_ind}
\vspace{-0.1cm}    
\end{align}
by using the triangle inequality. Now, we consider each expression separately. The first expression in \eqref{eq:k_s_ind}\off{, $\mathbb{V}(\tilde{p}_{\underline{Z}_{\mathcal{L}}|\underline{M}_{\mathcal{K}_s}=\underline{m}_{k_s}},\tilde{p}_{\underline{Z}_{\mathcal{W}}})$,} is bounded using \cite[Telescoping Expansion]{korada2010polar}, thus we have 
\vspace{-0.15cm}
\begin{align*}
\mathbb{V}\left(\tilde{p}_{\underline{Z}_{\mathcal{W}}|\underline{M}_{\mathcal{K}_s}=\underline{m}_{k_s}},\tilde{p}_{\underline{Z}_{\mathcal{W}}}\right) \leq \sum_{j=1}^{\tilde{n}} \mathbb{V}\left(\tilde{p}_{\underline{Z}_{\mathcal{W}}^{(j)}|\underline{M}_{\mathcal{K}_s}^{(j)}=\underline{m}_{k_s}^{(j)}},\tilde{p}_{\underline{Z}_{\mathcal{W}}}^{(j)}\right).
\vspace{-0.1cm}
\end{align*}
By applying \cite[Theorem 1]{SMSM}, it can be shown that for every $1 \leq j \leq \tilde{n}$, $\mathbb{V}(\tilde{p}_{\underline{Z}_{\mathcal{W}}^{(j)}|\underline{M}_{\mathcal{K}_s}^{(j)}=\underline{m}_{k_s}^{(j)}},\tilde{p}_{\underline{Z}_{\mathcal{W}}}^{(j)}) \leq \ell^{-\frac{t}{2}}$ for some $t \geq 1$ s.t. $\ell\epsilon =  \lceil t\log{\ell}\rceil$. 
We note, that as $t$ grows, the number $k_s$ of IS messages decreases. However, the information leakage decreases significantly as well. Now, we bound the third expression in \eqref{eq:k_s_ind}. As demonstrated in Appendix~\ref{appendix:IT-secrecy}, the same bound applies to the second expression in \eqref{eq:k_s_ind}. 
By using the independence between the messages, and applying the law of total probability to the distribution of IT-Eve's observations it can be shown that 
\begin{equation}\label{eq:bound_z}
\begin{aligned}
   \mathbb{V}\left(\tilde{p}_{\underline{Z}_{\mathcal{W}}},p_{\underline{Z}_{\mathcal{W}}}\right) \leq \mathbb{V}\left(p_{\underline{M}_{\mathcal{L}}},p_{U_{\underline{M}_{\mathcal{L}}}}\right). 
\end{aligned}
\end{equation}
We denote the KL-divergence between two distributions by $\mathbb{D}(\cdot||\cdot)$. Since the entropy of each bit in $\underline{M}_{\mathcal{L}}$ is at least $1-\delta_n$, it can be shown that $\mathbb{D}(p_{\underline{M}_{\mathcal{L}}}||p_{U_{\underline{M}_{\mathcal{L}}}}) \leq \ell \tilde{n} 2^{-n^\beta}$. By invoking the Pinsker inequality \cite{1053968}, we have that \eqref{eq:bound_z} is bounded by $\sqrt{2 \ell \tilde{n} 2^{-n^\beta}}$. Now, by using the triangle inequality and the fact that $\underline{V}_{\mathcal{K}_s} \rightarrow \underline{M}_{\mathcal{K}_s} \rightarrow \underline{Z}_{\mathcal{W}}$ is a Markov chain, we finally have: $\mathbb{V}(p_{\underline{Z}_{\mathcal{W}}|{\underline{V}_{\mathcal{K}_s}=\underline{v}_{k_s}}},p_{\underline{Z}_{\mathcal{W}}}) \leq 2\sqrt{2\ell \tilde{n} 2^{-n^\beta}} + \tilde{n} \ell^{-\frac{t}{2}}$. \off{Thus, for the information leakage to be negligible, we give lower and upper bounds on the size of $\ell$ compared to $n$.}The lower bound on $\ell$ is given by $\omega(\tilde{n}^{\frac{2}{t}})$, s.t. $\tilde{n} \ell^{-\frac{t}{2}}$ is negligible. The upper bound on $\ell$ is given by $o\left(2^{n^\beta}/\tilde{n}\right)$, s.t. $2\sqrt{2\ell \tilde{n} 2^{-n^\beta}}$ is negligible.

\section{Partial Encryption against Crypt-Eve \\ (Security  Proof Sketch of Theorem~\ref{Individual-SS-CCA1})}\label{sec:Crypto-Eve}
In this section, we give a security proof sketch of Theorem~\ref{Individual-SS-CCA1}. Due to space limitations, the full proof is provided in  Appendix~\ref{appendix:individual-ss-cca1}. The proof is based on the equivalence between semantic security and indistinguishability \cite{goldwasser2019probabilistic}. 
Considering the maximal advantage for Crypto-Eve (maximal divination from uniform distribution probability), given the almost uniform messages after the source encoding stage (see Section~\ref{sec:main_results}), we prove that each column of the message matrix $\underline{M}_{\mathcal{L}}$ is IIND-CCA1 as given in \cite[Definition 4]{cohen2022partial}, and thus it is ISS-CCA1.\off{The proof follows similar techniques as used in \cite[Theorem 3]{cohen2022partial} with the primary distinction residing in the distribution of the input messages.}

For any column $1 \leq j \leq \tilde{n}$ of the message matrix, Crypto-Eve chooses $i^{*} \in [1,k_s]$ (the case for $i^{*} \in [k_s+1,\ell]$ follows analogously) and two messages $M_{i^{*},1}$ and $M_{i^{*},2}$. Crypto-Eve sends Alice a polynomial amount of ciphertexts and receives back the decrypted messages. Alice chooses uniformly at random $h \in \{1,2\}$ and draws $\{M_{i}\}_{i=1}^{\ell,i\neq i^{*}}$ from the distribution induced by the source messages. To show a stronger statement than in Definition~\ref{def:individuall-SS-CCA1}, we assume Alice provides Crypto-Eve with the bits in indexes $\{1,...,k_s\} \setminus i^{*}$, and show that Crypto-Eve can't distinguish between the bins $b_1=(M_1,...,M_{i^{*},1},...,M_{k_s})$ and $b_2 = (M_1,...,M_{i^{*},2},...,M_{k_s})$. Alice encrypts the message using $Crypt_2$ from \eqref{eq:encryption_crypt2}. \off{Alice encodes the message using the $k_s$-IS secured code from Appendix~\ref{appendix:msg_encoder} and then encrypts the codeword using $Crypt_1$ from \eqref{eq:encryption}.} Crypto-Eve observes $w$ bits of the encrypted codeword as plaintext and thus can reduce the number of possible codewords in each of the two bins. We denote the group of the remaining possible codewords in bins $b_1$ and $b_2$ by $\mathcal{B}_1$ and $\mathcal{B}_2$ respectively. In \cite[Section IV.B]{SMSM}, it was shown that for some $\epsilon' > 0$, in high probability $(1-\epsilon')2^{\ell\epsilon} \leq |\mathcal{B}_1|,|\mathcal{B}_2| \leq (1+\epsilon')2^{\ell\epsilon}$.

Now, we consider the best possible scenario for Crypto-Eve: 1) $|\mathcal{B}_1| \geq |\mathcal{B}_2|$, 2) The induced probability of the codewords from $\mathcal{B}_1$ is as high as possible while the induced probability of the codewords from $\mathcal{B}_2$ is as low as possible, and 3) The distinguishability from $Crypt_1$ is as high as possible and denoted by $\epsilon_{ss-cca1} \leq \frac{1}{c^d}$. 

We bound the induced probabilities of the codewords from $\mathcal{B}_1$ and $\mathcal{B}_2$ by $p_{max}$ and $p_{min}$ respectively. 
\off{We upper bound the induced probabilities of the codewords from $\mathcal{B}_1$ by $p_{max}$ and lower bound the probabilities of the codewords from $\mathcal{B}_2$ by $p_{min}$.} If $j \in \mathcal{H}_V$, it can be shown that $p_{max}$ and $p_{min}$ can be upper/lower bounded by $ \left(\left(1 \pm \sqrt{1-(1-\delta_n)^{\ln{4}}}\right)/2\right)^{k_w}$,  respectively \cite{topsoe2001bounds}, where $\delta_n$ is a parameter of the source coding scheme (see Section~\ref{sec:main_results}).
\off{We denote $\alpha \in \{\mathcal{B}_1 \cup \mathcal{B}_2\}$ as one of the possible codewords.} Thus, as demonstrated in Appendix~\ref{appendix:individual-ss-cca1}, Crypto-Eve's overall advantage is given by
\begin{align}
    \mathbb{P}[M_{i^{*},1}] - \frac{1}{2} &=  \frac{|\mathcal{B}_1| \cdot p_{max} - |\mathcal{B}_2| \cdot p_{min}}{|\mathcal{B}_1| \cdot p_{max} + |\mathcal{B}_2| \cdot p_{min}} + \epsilon_{ss-cca1} \nonumber \\
    & \leq 2^{\frac{-n^{\beta}}{2}} k_w + \ell^{-t} + \frac{1}{c^{d}}, \label{eq:SS-CCA1-sec-summ}
\end{align}
where $t \geq 1$ is s.t. $\ell\epsilon = \lceil t \log{\ell} \rceil$. For every $d'$, by taking $\ell$ large enough, $\ell \leq 2^{\frac{n^\beta}{2(t+1)}}$, an appropriate $d$ and $t = \log{\ell}$, the expression in \eqref{eq:SS-CCA1-sec-summ} is smaller than $\frac{1}{c^{d'}}$. 
The case for $j \in \mathcal{J}_V$ is trivial and directly follows from the SS-CCA1 of $Crpyt_1$, which is used to encrypt the uniform seed. Thus, each column $1 \leq j \leq \tilde{n}$ is IIND-CCA1 and equivalently ISS-CCA1.

\section{Conclusions And Future Work} \label{sec:discussion}
In this paper, we introduce a novel PQ individual SS-CCA1 cryptosystem for non-uniform messages at high data rates. This result is given by using a polar source coding scheme with a sub-linear shared seed, a secure channel coding scheme that mixes the almost uniform information, and encrypting a small amount of the mixed data and the seed by a PQ cryptosystem.

Future works include mixing the data by secure linear coding schemes (e.g., as given in \cite{bhattad2005weakly,silva2009universal,SMSM}) for computationally efficient decoding \cite{silva2009fast}, and a converse analysis for the individual secure source and channel scheme considered. Extensions also include the study of secure linear channel coding schemes for the proposed PQ cryptosystem to guarantee reliable communications over noisy channels \cite{koetter2008coding,silva2008rank,silva2011universal}.




\bibliographystyle{IEEEtran}
\bibliography{refs}

\appendices
\section{Source and IS Channel Coding Schemes} 
For completeness, first, we provide here the source coding construction as presented in \cite[Proposition 4]{NegligbleCost} and the IS random channel coding construction as presented in \cite[Section \uppercase\expandafter{\romannumeral4}]{SMSM}, we consider for the proposed NU-HUNCC scheme. Then, we analyze the reliability of NU-HUNCC coding scheme suggested in Section~\ref{sec:main_results}, using the source and IS channel coding schemes on $\ell$ non-uniform messages of size $n$.

\subsection{Polar Source Code}\label{appendix:src_code}
Let $m \in \mathbb{N}$ and the blocklength $n = 2^m$. We denote by $\underline{G}_n$, the polarization transform as defined in \cite{ArikanBase2009} s.t. $\underline{G}_n = \underline{P}_n\begin{bmatrix} 1 & 0 \\ 1 & 1 \end{bmatrix}^{\otimes m}$, where $\otimes$ is the Kronecker product, and $\underline{P}_n$ is the bit reversal matrix. In our proposed scheme, each row of the source message matrix $\underline{V}_{\mathcal{L}}$ is separately encoded by the polarization transformation $\underline{G}_n$: $\underline{A}_n = 
 \underline{V}_\mathcal{L} \cdot \underline{G}_n$. Let $\delta_n \triangleq 2^{-n^\beta}$, for $\beta \in [0,\frac{1}{2})$. For each row $A \subset \underline{A}_n$, the bits are divided into three groups:
\begin{equation} \label{eq:unreliable-group}
    \begin{aligned}
        1) \quad \mathcal{H}_V \triangleq \left\{ j \in [1,n]: H\left(A^{(j)}|A^{j-1}\right) > 1 - \delta_n \right\},
    \end{aligned}
\end{equation}
\begin{equation*}
    \begin{aligned}
     \hspace{-1.05cm}   2) \quad \mathcal{U}_V \triangleq \left\{ j \in [1,n]: H\left(A^{(j)}|A^{j-1}\right) < \delta_n \right\},
    \end{aligned}
\end{equation*}
\begin{equation*} \label{eq:seed-group}
    \begin{aligned}
     \hspace{-4.00cm}  3) \quad  \mathcal{J}_V \triangleq (\mathcal{U}_V \cup \mathcal{H}_V)^{C}.
    \end{aligned}
\end{equation*}
Traditionally, the compressed message in polar codes-based source coding is obtained from the concatenation of groups, $\mathcal{H}_V$, and $\mathcal{J}_V$, the bits with non-negligible entropy. The entropy of the bits from group $\mathcal{J}_V$ is not necessarily high, and they are not uniform.
To ensure the uniformity of those bits, they undergo one-time padding with the uniform seed. We denote by $U_{d_{J}}$, the uniform seed of size $|\mathcal{J}_{V}| \triangleq d_{J}$. A different seed is used for each row, thus we denote the seed matrix by $\underline{U}_{d_{J}} \in \mathbb{F}_{2}^{\ell \times d_{J}}$. Finally, the $i$-th row of the message matrix is given by
\begin{equation} \label{eq:output_src}
    \begin{aligned}
        \underline{M}_{\mathcal{L},i} \triangleq \left[A_i[\mathcal{H}_V],A_i[\mathcal{J}_V] \oplus \underline{U}_{d_{J},i}\right].
    \end{aligned}
\end{equation}
Here $\oplus$ denotes the xor operation over $\mathbb{F}_2$.  
The size of each almost uniform message obtained at the outcome of the source coding scheme is $\tilde{n} = |\mathcal{H}_V| + d_{J}$.

The decoding process is divided into two parts. First, Bob one-time pads the bits from the group $\mathcal{J}_V$ using the seed shared with him by Alice. Second, Bob uses a successive cancellation decoder to reliably decode the original message \cite{ArikanBase2009,cronie2010lossless}. In the suggested scheme, this decoding process is performed separately on each row of the message matrix.

\subsection{IS Random Channel Coding}\label{appendix:msg_encoder}
The IS random channel coding scheme is applied separately on each column of the message matrix $\underline{M}_{\mathcal{L}}$. Let $w < \ell$ be the number of links observed by IT-Eve, and $k_s \leq \ell - w - \ell\epsilon$ be the number of messages kept secured from IT-Eve s.t. $\ell\epsilon$ is an integer.\off{At the end of the secrecy proof, we discuss the effect $\ell\epsilon$ has on the information leakage to IT-Eve.} We denote $k_w \triangleq w + \ell\epsilon$. Each column of the message matrix is divided into two parts of sizes $k_s$ and $k_w$. The first part is denoted by $M_{k_s} \in \mathbb{F}_2^{k_s}$, and the second is denoted by $M_{k_w} \in \mathbb{F}_2^{k_w}$. We now give the detailed code construction for the IS random channel code applied by Alice on each $j$-th column.

\textit{\underline{Codebook Generation}}: Let $P(x) \sim Bernouli(\frac{1}{2})$. There are $2^{k_s}$ possible messages for $M_{k_s}$ and $2^{k_w}$ possible messages for $M_{k_w}$. For each possible message $M_{k_s}$, generate $2^{k_w}$ independent codewords $x^{\ell}(e)$, $1 \leq e \leq 2^{k_w}$, using the distribution $P(X^{\ell}) = \prod_{j=1}^{\ell}P(X_j)$. Thus, we have $2^{k_s}$ bins, each having $2^{k_w}$ possible codewords. The length of the codeword remains the same as the length of the message since the links are noiseless \cite{SMSM} and error correction is not required.\off{, e.g., as given in \cite{cohen2022partial}.}

\textit{\underline{Encoding}}: For each column in the message matrix, $k_s$ bits are used to choose the bin, while the remaining $k_w$ bits are used to select the codeword inside the bin.

\textit{\underline{Decoding}}: Bob decodes each column separately. \off{by finding the received codeword of each column in the codebook.} If the codeword appears only once in the codebook, then $M_{k_s}$ is the index of the bin in which the codeword was found, and $M_{k_w}$ is the index of the codeword inside the bin. However, if the codeword appears more than once, a decoding error occurs. The probability of a codeword appearing more than once in a codebook exponentially decreases as a function of $\ell$. The reliability analysis of the code is given in \cite[Section \uppercase\expandafter{\romannumeral4}]{SMSM}.

\subsection{Reliability} \label{appendix:reliability}
The reliability of the proposed end-to-end source-channel coding schemes is a consequence of the reliability of the source and channel decoders over the $\ell$ rows of the source matrix and $\tilde{n}$ columns of the message matrix. Hence, the proof is based on a similar technique given in \cite{NegligbleCost}, with the additional use of a union bound analyzed herein. \off{We denote the source encoder and decoder pair for each row by $f_{s,n}$ and $g_{s,n}$, respectively, and the channel encoder and decoder for each column by $f_{c,\ell}$ and $g_{c,\ell}$, respectively.}

First, each column of the encoded message matrix is separately decoded. We bound the decoding error probability of the channel coding scheme. We denote the uniformly distributed message matrix by $\underline{\tilde{M}}_{\mathcal{L}}$ and its corresponding uniform distribution by $p_{U_{\underline{M}_{\mathcal{L}}}}$. We consider the optimal coupling as given in \cite[Lemma 3.6]{aldous1983random}, between $\underline{M}_{\mathcal{L}}$ and $\underline{\tilde{M}}_{\mathcal{L}}$. We denote $\varepsilon \triangleq \{\underline{M}_{\mathcal{L}} \neq \underline{\tilde{M}}_{\mathcal{L}}\}$. From the coupling lemma, we have that $\mathbb{P}[\varepsilon] = \mathbb{V}(p_{\underline{M}_{\mathcal{L}}},p_{U_{\underline{M}_{\mathcal{L}}}})$. The decoded message matrix is denoted by $\underline{\hat{M}}_{\mathcal{L}}$. Thus, the decoding error probability of the message matrix is given by
\begin{align}
    &\mathbb{P}\left[\underline{\hat{M}}_{\mathcal{L}} \neq \underline{M}_{\mathcal{L}} | \underline{M}_{\mathcal{L}} = \underline{m}_{\ell}\right] \nonumber \\
    & \quad= \mathbb{P}\left[\underline{\hat{M}}_{\mathcal{L}} \neq \underline{M}_{\mathcal{L}} | \underline{M}_{\mathcal{L}} = \underline{m}_{\ell},\varepsilon^C\right]\mathbb{P}\left[\varepsilon^C\right] \nonumber \\
    & \quad\quad + \mathbb{P}\left[\underline{\hat{M}}_{\mathcal{L}} \neq  \underline{M}_{\mathcal{L}} | \underline{M}_{\mathcal{L}} = \underline{m}_{\ell},\varepsilon\right]\mathbb{P}\left[\varepsilon\right] \nonumber \\
    & \quad \leq \mathbb{P}\left[\underline{\hat{M}}_{\mathcal{L}} \neq  \underline{M}_{\mathcal{L}} | \underline{M}_{\mathcal{L}} = \underline{m}_{\ell},\varepsilon^C\right] + \mathbb{P}\left[\varepsilon\right] \nonumber \\
    & \quad \overset{(a)}{=} \mathbb{P}\left[\underline{\hat{M}}_{\mathcal{L}} \neq  \underline{M}_{\mathcal{L}} | \underline{M}_{\mathcal{L}} = \underline{m}_{\ell},\varepsilon^C\right] + \mathbb{V}(p_{\underline{M}_{\mathcal{L}}},p_{U_{\underline{M}_{\mathcal{L}}}}) \nonumber \\
    & \quad \overset{(b)}{\leq} \mathbb{P}\left[\underline{\hat{\tilde{M}}}_{\mathcal{L}} \neq \underline{\tilde{M}}_{\mathcal{L}} | \underline{\tilde{M}}_{\mathcal{L}} = \underline{m}_{\ell}\right] + \sqrt{2 \tilde{n} \ell 2^{-n^\beta}}, \label{eq:column_Pe}
\end{align}
where (a) is from the coupling Lemma \cite[Lemma 3.6]{aldous1983random}, and (b) is from \eqref{eq:uniform-bound}. The left term in \eqref{eq:column_Pe} refers to the decoding error probability of the message matrix assuming the message distribution was entirely uniform. The decoding error probability of each column is upper bounded by $O\left(2^{-\ell}\right)$ \cite[Section \uppercase\expandafter{\romannumeral4}]{SMSM}. Thus, the error probability for the entire message matrix is given by the union bound on all the columns of the message matrix
\begin{equation}\label{eq:Column_Pe}
    \begin{aligned}
    \mathbb{P}\left[\underline{\hat{\tilde{M}}}_{\mathcal{L}} \neq \underline{\tilde{M}}_{\mathcal{L}} | \underline{\tilde{M}}_{\mathcal{L}} = \underline{m}_{\ell}\right]  \leq  \tilde{n} \cdot O\left(2^{-\ell}\right).
    \end{aligned}
\end{equation}
After obtaining the decoded message matrix, Bob can decode each row of the matrix separately to recover the source messages. Since Bob has the shared seed, he first one-time pads the bits from group $\mathcal{J}_V$ as described in Appendix~\ref{appendix:src_code}. Bob continues by employing successive cancellation decoding \cite{cronie2010lossless}\cite{arikan2009rate}. Such that, the error probability for the $i$-th row is given by
\begin{equation} \label{eq:single_row_pe}
\mathbb{P}\left[\underline{\hat{V}}_{\mathcal{L},i} \neq g_{s,n}\left(f_{s,n}(\underline{V}_{\mathcal{L},i},\underline{U}_{d_J,i},\underline{U}_{d_J,i})\right)\right] \leq 2^{-n^{\beta}},
\end{equation}
for $\beta \in [0,\frac{1}{2})$ as have been chosen in Appendix~\ref{appendix:src_code}. 

Now, using the union bound on all the rows in the recovered message matrix in \eqref{eq:single_row_pe}, we have
\begin{equation} \label{eq:UnionError}
    \mathbb{P}\left(\bigcup_{i=1}^{\ell} \left\{\underline{\hat{V}}_{\mathcal{L},i} \neq g_{s,n}\left(f_{s,n}(\underline{V}_{\mathcal{L},i},\underline{U}_{d_J,i},\underline{U}_{d_J,i})\right)\right\}\right) \leq \ell2^{-n^{\beta}}.
\end{equation}

Finally, we bound the total error probability for the source and channel coding schemed together obtaining
\begin{equation*}\label{eq:tot-Pe}
\begin{aligned}
    &\mathbb{P}\left[V_{\mathcal{L}} \neq \underline{\hat{V}}_{\mathcal{L}}\right] \\
    & \quad \leq \mathbb{P}\left[V_{\mathcal{L}} \neq \underline{\hat{V}}_{\mathcal{L}}|\underline{M}_{\mathcal{L}} \neq \underline{\hat{M}}_{\mathcal{L}}\right]\mathbb{P}\left[\underline{M}_{\mathcal{L}} \neq \underline{\hat{M}}_{\mathcal{L}}\right] \\
    & \quad\quad\quad + \mathbb{P}\left[V_{\mathcal{L}} \neq \underline{\hat{V}}_{\mathcal{L}}|\underline{M}_{\mathcal{L}} = \underline{\hat{M}}_{\mathcal{L}}\right]\mathbb{P}\left[\underline{M}_{\mathcal{L}} = \underline{\hat{M}}_{\mathcal{L}}\right] \\
    & \quad \leq  \mathbb{P}\left[\underline{M}_{\mathcal{L}} \neq \underline{\hat{M}}_{\mathcal{L}}\right] \\
    & \quad\quad\quad + \mathbb{P}\left[V_{\mathcal{L}} \neq \underline{\hat{V}}_{\mathcal{L}}|\underline{M}_{\mathcal{L}} = \underline{\hat{M}}_{\mathcal{L}}\right] \\
    & \quad \overset{(a)}{\leq} \tilde{n} O\left(2^{-\ell}\right) +  \sqrt{2 \tilde{n}  \ell  2^{-n^\beta}} + \ell2^{-n^{\beta}},
\end{aligned}
\end{equation*}
where (a) is given by \eqref{eq:column_Pe}, \eqref{eq:Column_Pe}, and \eqref{eq:UnionError}. 

By requiring $\ell$ to be lower bounded by $\omega(\log{\tilde{n}})$, the expression $\tilde{n} O\left(2^{-\ell}\right)$ becomes negligible. In addition, requiring $\ell$ to be upper bounded by $ o\left(2^{n^\beta}/\tilde{n}\right)$, the expression $\sqrt{2 \tilde{n} \ell 2^{-n^\beta}} + \ell2^{-n^{\beta}}$ becomes negligible as well. By choosing $\ell$ that upholds both bounds the decoding error probability becomes negligible.
\off{For the error probability to be negligible we give $\ell$ upper and lower bounds. The lower bound of $\ell$ is required for $\tilde{n} O\left(2^{-\ell}\right)$ to be negligible. Thus, we require that $\ell = \omega(\log{\tilde{n}})$. The upper bound is required for the expression $\sqrt{2 \tilde{n} \ell 2^{-n^\beta}} + \ell2^{-n^{\beta}}$ to be negligible. This is obtained by choosing $\ell$ s.t. $\ell = o\left(2^{n^\beta}/\tilde{n}\right)$. By choosing $\ell$ that upholds both bounds the decoding error probability becomes negligible.}

\section{$k_s$-IS Information Leakage against IT-Eve}\label{appendix:IT-secrecy}
We give here the full secrecy analysis of the $k_s$-IS coding scheme (security proof of Theorem~\ref{Direct}). We assume a noiseless multipath communication system with $\ell$ links and an eavesdropper, IT-Eve, with access to any subset $\mathcal{W} \subset \mathcal{L}$ of the links s.t. $|\mathcal{W}| \triangleq w < \ell$. Alice wants to send $\ell$ confidential messages to Bob while keeping IT-Eve ignorant about any set of $k_s \leq \ell - w - \ell\epsilon$ messages individually. We assume Alice uses the source coding scheme from Appendix~\ref{appendix:src_code} and the channel coding scheme from Appendix~\ref{appendix:msg_encoder}.

Alice encodes the source message matrix $\underline{V}_{\mathcal{L}}$ using the polar codes-based source encoder (Appendix~\ref{appendix:src_code}). The almost uniform message at the output of the source encoder is of size $\tilde{n} = |\mathcal{H}_V| + d_{J}$, where $d_{J} = |\mathcal{J}_V|$. Thus, we denote the output of the source encoder as given in \eqref{eq:output_src} by $\underline{M}_\mathcal{L} \in \mathbb{F}_2^{\ell \times \tilde{n}}$.

We denote the set $\mathcal{K}_s \subset \mathcal{L}$ s.t. $|\mathcal{K}_s| \triangleq k_s$, and the set  $\mathcal{K}_w \triangleq \mathcal{L} \setminus \mathcal{K}_s$. We start by showing that the channel scheme is $k_s$-IS. We denote by $\underline{M}_{\mathcal{K}_s} \subset \underline{M}_{\mathcal{L}}$ the subset of the secured messages and by $\underline{M}_{\mathcal{K}_w} \subset \underline{M}_{\mathcal{L}} \setminus \underline{M}_{\mathcal{K}_s}$ the rest of the messages. The distribution of $\underline{Z}_{\mathcal{W}}$ induced by the uniform message matrix is denoted by $\tilde{p}_{\underline{Z}_{\mathcal{W}}}$ or $\tilde{p}_{\underline{Z}_{\mathcal{W}}|\underline{M}_{\mathcal{K}_s}=\underline{m}_{k_s}}$. For any $\underline{m}_{k_s} \in \underline{\mathcal{M}}_{k_s}$
\begin{alignat}{1}
    &\mathbb{V}\left(p_{\underline{Z}_{\mathcal{W}}|\underline{M}_{\mathcal{K}_s}=\underline{m}_{k_s}},p_{\underline{Z}_{\mathcal{W}}}\right) \label{FullSecrecy}\\
    & \quad \overset{(a)}{\leq} \mathbb{V}\left(p_{\underline{Z}_{\mathcal{W}}|\underline{M}_{\mathcal{K}_s}=\underline{m}_{k_s}},\tilde{p}_{\underline{Z}_{\mathcal{W}}|\underline{M}_{\mathcal{K}_s}=\underline{m}_{k_s}}\right) \label{eq:CondSec} \\
    & \quad\quad + \mathbb{V}\left(\tilde{p}_{\underline{Z}_{\mathcal{W}}|\underline{M}_{\mathcal{K}_s}=\underline{m}_{k_s}},\tilde{p}_{\underline{Z}_{\mathcal{W}}}\right) \label{eq:UnifSec} \\
    & \quad\quad + \mathbb{V}\left(\tilde{p}_{\underline{Z}_{\mathcal{W}}},p_{\underline{Z}_{\mathcal{W}}}\right), \label{eq:FullZ}
\end{alignat}
where (a) is from the triangle inequality. Since this is true for all  $\underline{m}_{k_s} \in \underline{\mathcal{M}}_{k_s}$, from now on we omit the equality $\underline{M}_{\mathcal{K}_s} = \underline{m}_{k_s}$ for ease of notation.

Now, we bound each of the expressions \eqref{eq:CondSec}-\eqref{eq:FullZ}, starting with \eqref{eq:CondSec}
\begin{align*}
    & \mathbb{V}\left(p_{\underline{Z}_{\mathcal{W}}|\underline{M}_{\mathcal{K}_s}},\tilde{p}_{\underline{Z}_{\mathcal{W}}|\underline{M}_{\mathcal{K}_s}}\right) \\
    & \quad = \sum_{\underline{z}_{w}} \left|p_{\underline{Z}_{\mathcal{W}}|\underline{M}_{\mathcal{K}_s}}(\underline{z}_{w}|\underline{m}_{k_s})-\tilde{p}_{\underline{Z}_{\mathcal{W}}|\underline{M}_{\mathcal{K}_s}}(\underline{z}_{w}|\underline{m}_{k_s})\right| \\
    & \quad = \sum_{\underline{z}_{w}}\left|\sum_{\underline{m}_{k_w}}\left(p_{\underline{Z}_{\mathcal{W}}\underline{M}_{\mathcal{K}_w}|\underline{M}_{\mathcal{K}_s}}(\underline{z}_{w},\underline{m}_{k_w}|\underline{m}_{k_s}) \right. \right. \\
    & \quad\quad\quad\quad\quad\quad\quad \left. \left. - \tilde{p}_{\underline{Z}_{\mathcal{W}}\underline{M}_{\mathcal{K}_w}|\underline{M}_{\mathcal{K}_s}}(\underline{z}_{w},\underline{m}_{k_w}|\underline{m}_{k_s})\right)\right| \\
    & \quad = \sum_{\underline{z}_{w}}\left|\sum_{\underline{m}_{k_w}}\left(p_{\underline{Z}_{\mathcal{W}}|\underline{M}_{\mathcal{K}_w}\underline{M}_{\mathcal{K}_s}}(\underline{z}_{w}|\underline{m}_{k_w},\underline{m}_{k_s}) \right. \right. \\
    & \quad\quad\quad\quad\quad\quad\quad\quad\quad\quad\quad \left. \left.\cdot p_{\underline{M}_{\mathcal{K}_w}|\underline{M}_{\mathcal{K}_s}}(\underline{m}_{k_w}|\underline{m}_{k_s}) \right. \right. \\
    & \quad\quad\quad\quad\quad \left. \left. -p_{\underline{Z}_{\mathcal{W}}|\underline{M}_{\mathcal{K}_w}\underline{M}_{\mathcal{K}_s}}(\underline{z}_{w}|\underline{m}_{k_w},\underline{m}_{k_s}) \right. \right. \\
    & \quad\quad\quad\quad\quad\quad\quad\quad\quad\quad\quad \left. \left. \cdot \tilde{p}_{\underline{M}_{\mathcal{K}_w}|\underline{M}_{\mathcal{K}_s}}(\underline{m}_{k_w}|\underline{m}_{k_s})\right)\right| \\
    & \quad = \sum_{\underline{z}_\mathcal{W}} p_{\underline{Z}_{\mathcal{W}}|\underline{M}_{\mathcal{K}_w}\underline{M}_{\mathcal{K}_s}}(\underline{z}_{w}|\underline{m}_{k_w},\underline{m}_{k_s}) \cdot \\ 
    & \quad\quad\quad\quad \left|\sum_{\underline{m}_{k_w}} \left(p_{\underline{M}_{\mathcal{K}_w}|\underline{M}_{\mathcal{K}_s}}(\underline{m}_{k_w}|\underline{m}_{k_s}) \right. \right. \\
    &  \quad\quad\quad\quad\quad\quad\quad\quad\quad \left. \left. -\tilde{p}_{\underline{M}_{\mathcal{K}_w}|\underline{M}_{\mathcal{K}_s}}(\underline{m}_{k_w}|\underline{m}_{k_s})\right)\right| \\
    & \quad \overset{(a)}{\leq} \sum_{\underline{z}_\mathcal{W}} \sum_{\underline{m}_{k_w}} p_{\underline{Z}_{\mathcal{W}}|\underline{M}_{\mathcal{K}_w}\underline{M}_{\mathcal{K}_s}}(\underline{z}_{w}|\underline{m}_{k_w},\underline{m}_{k_s}) \cdot \\
    & \quad\quad\quad \left|p_{\underline{M}_{\mathcal{K}_w}|\underline{M}_{\mathcal{K}_s}}(\underline{m}_{k_w}|\underline{m}_{k_s})-\tilde{p}_{\underline{M}_{\mathcal{K}_w}|\underline{M}_{\mathcal{K}_s}}(\underline{m}_{k_w}|\underline{m}_{k_s})\right| \\
    & \quad \overset{(b)}{=} \sum_{\underline{m}_{k_w}} \left|p_{\underline{M}_{\mathcal{K}_w}}(\underline{m}_{k_w}) - \tilde{p}_{\underline{M}_{\mathcal{K}_w}}(\underline{m}_{k_w})\right| \\
    & \quad = \mathbb{V}\left(p_{\underline{M}_{\mathcal{K}_w}},p_{U_{\underline{M}_{\mathcal{K}_w}}}\right) \leq \mathbb{V}\left(p_{\underline{M}_{\mathcal{L}}},p_{U_{\underline{M}_{\mathcal{L}}}}\right),
\end{align*}
where (a) holds from the triangle inequality, (b) holds from the independence between messages and $p_{U_{\underline{M}_{\mathcal{K}_w}}}$ is the uniform distribution of the uniform matrix $\underline{\tilde{M}}_{\mathcal{K}_w}$. We continue with bounding \eqref{eq:FullZ}
\begin{multline*}
    \begin{aligned}
    &\mathbb{V}\left(\tilde{p}_{\underline{Z}_{\mathcal{W}}},p_{\underline{Z}_{\mathcal{W}}}\right) \\
       & \quad= \sum_{\underline{z}_{w}}\left|\tilde{p}_{\underline{Z}_{\mathcal{W}}}(\underline{z}_{w}) - p_{\underline{Z}_{\mathcal{W}}}(\underline{z}_{w})\right| \\
       & \quad = \sum_{\underline{z}_{w}}\left|\sum_{\underline{m}_{k_w},\underline{m}_{k_s}}\left(\tilde{p}_{\underline{Z}_{\mathcal{W}} \underline{M}_{\mathcal{K}_w} \underline{M}_{\mathcal{K}_s}}(\underline{z}_{w},\underline{m}_{k_w},\underline{m}_{k_s})  \right. \right. \\ 
       & \quad\quad\quad\quad\quad\quad\quad\quad\quad \left. \left. -p_{\underline{Z}_{\mathcal{W}} \underline{M}_{\mathcal{K}_w} \underline{M}_{\mathcal{K}_s}}(\underline{z}_{w},\underline{m}_{k_w},\underline{m}_{k_s})\right)\right| \\
       & \quad = \sum_{\underline{z}_{w}}\left|\sum_{\underline{m}_{\ell}}\left(p_{\underline{Z}_{\mathcal{W}}|\underline{M}_{\mathcal{L}}}(\underline{z}_{w}|\underline{m}_{\ell})  \cdot p_{U_{\underline{M}_{\mathcal{L}}}}(\underline{m}_{\ell}) \right. \right. \\
       & \quad\quad\quad\quad\quad\quad \left. \left.- p_{\underline{Z}_{\mathcal{W}} | \underline{M}_{\mathcal{L}}}(\underline{z}_{w}|\underline{m}_{\ell}) \cdot p_{\underline{M}_{\mathcal{L}}}(\underline{m}_{\ell})\right)\right| \\
       & \quad \overset{(a)}{\leq} \sum_{\underline{m}_{\ell}}\sum_{\underline{z}_{w}} p_{\underline{Z}_{\mathcal{W}}|\underline{M}_{\mathcal{L}}}(\underline{z}_{w}|\underline{m}_{\ell}) \cdot \left| p_{U_{\underline{M}_{\mathcal{L}}}}(\underline{m}_{\ell}) - p_{M_{\mathcal{L}}}(\underline{m}_{\ell})\right| \\
       & \quad = \sum_{\underline{m}_{\ell}} \left|p_{U_{\underline{M}_{\mathcal{L}}}}(\underline{m}_{\ell}) - p_{\underline{M}_{\mathcal{L}}}(\underline{m}_{\ell})\right|  = \mathbb{V}\left(p_{\underline{M}_{\mathcal{L}}},p_{U_{\underline{M}_{\mathcal{L}}}}\right),
    \end{aligned}
\end{multline*}
where inequality (a) follows from the triangle inequality.

We now bound $\mathbb{D}(p_{\underline{M}_{\mathcal{L}}}||p_{U_{\underline{M}_{\mathcal{L}}}})$\off{. For ease of notation, we denote a column from $\underline{M}_{\mathcal{L}}$ by $M(j)$} and obtain 
\begin{equation*}
\begin{aligned}
    &\mathbb{D}\left(p_{\underline{M}_{\mathcal{L}}}||p_{U_{\underline{M}_{\mathcal{L}}}}\right) \\
    & \quad = \sum_{\underline{m}_{\ell}} p_{\underline{M}_{\mathcal{L}}}(\underline{m}_{\ell}) \cdot \log_{2} \left(\frac{p_{\underline{M}_{\mathcal{L}}}(\underline{m}_{\ell})}{p_{U_{\underline{M}_{\mathcal{L}}}}(\underline{m}_{\ell})}\right) \\
    & \quad = \log_{2} (2^{\ell \cdot \tilde{n}}) - H\left(\underline{M}_{\mathcal{L}}\right) \\
    & \quad = \ell \tilde{n} - H\left(\underline{M}_{\mathcal{L}}\right) \\
    & \quad \overset{(a)}{=} \ell \tilde{n} - \ell \cdot \sum_{j=1}^{\tilde{n}}  H\left(M^{(j)}|M^{j-1}\right) \\
    & \quad \overset{(b)}{\leq} \ell \tilde{n} - \ell  \tilde{n} (1-\delta_n) = \ell  \tilde{n}  \delta_n \leq \ell  \tilde{n} 2^{-n^\beta},
\end{aligned}
\end{equation*}
where (a) holds from the independence between messages, and (b) is given using the source scheme in Appendix~\ref{appendix:src_code}.

Now, we invoke the Pinsker inequality \cite{1053968} to bound the variational distance between the distribution $p_{\underline{M}_{\mathcal{L}}}$ and the uniform distribution $p_{U_{\underline{M}_{\mathcal{L}}}}$. Thus, we have
\begin{equation*}
    \begin{aligned}
        \frac{1}{2}\mathbb{V}^2\left(p_{\underline{M}_{\mathcal{L}}},p_{U_{\underline{M}_{\mathcal{L}}}}\right) \leq  \mathbb{D}\left(p_{\underline{M}_{\mathcal{L}}}||p_{U_{\underline{M}_{\mathcal{L}}}}\right) \leq \ell \tilde{n} 2^{-n^\beta},
    \end{aligned}
\end{equation*}
s.t. we bound the expressions in \eqref{eq:CondSec} and \eqref{eq:FullZ} by
\begin{equation} \label{eq:uniform-bound}
    \begin{aligned}
        \mathbb{V}\left(p_{\underline{M}_{\mathcal{L}}},p_{U_{\underline{M}_{\mathcal{L}}}}\right) \leq \sqrt{2 \ell \tilde{n} 2^{-n^\beta}}.
    \end{aligned}
\end{equation}

\off{To bound \eqref{eq:UnifSec}, we first present the following lemma.
\begin{lemma} \label{Lemma:HelpLemma}
Let $a_i,b_i$, $1 \leq i \leq n$ be a series of non-negative numbers. Then, the following inequality holds 
\begin{gather*}
    \left|\prod_{i=1}^{n}a_i - \prod_{i=1}^{n}b_i\right| \leq \sum_{i=1}^{n}|a_i-b_i|\cdot \prod_{j=1}^{i-1}a_j \cdot \prod_{k=i+1}^{n}b_k.
\end{gather*}
\end{lemma}
Using Lemma \ref{Lemma:HelpLemma}, we have}
We now bound \eqref{eq:UnifSec} and obtain
\begin{equation} \label{eq:HelpLemmaBound}
\begin{aligned}
    &\mathbb{V}\left(\tilde{p}_{\underline{Z}_{\mathcal{W}}|\underline{M}_{\mathcal{K}_s}},\tilde{p}_{\underline{Z}_{\mathcal{W}}}\right) \\
    & \quad = \sum_{\underline{z}_w}\left|\tilde{p}_{\underline{Z}_{\mathcal{W}}|\underline{M}_{\mathcal{K}_s}}(\underline{z}_w|\underline{m}_{k_s})-\tilde{p}_{\underline{Z}_{\mathcal{W}}}(\underline{z}_w)\right| \\
    & \quad \overset{(a)}{=} \sum_{\underline{z}_\mathcal{W}}\left|\prod_{j=1}^{\tilde{n}}\tilde{p}_{\underline{Z}_{\mathcal{W}}^{(j)}|\underline{M}_{\mathcal{K}_s}^{(j)}}(\underline{z}^{(j)}|\underline{m}_{k_s}^{(j)}) - \prod_{j=1}^{\tilde{n}}\tilde{p}_{\underline{Z}_{\mathcal{W}}^{(j)}}(\underline{z}^{(j)})\right| \\
    & \quad \overset{(b)}{\leq} \sum_{\underline{z}_\mathcal{W}} \sum_{j=1}^{\tilde{n}} \left|\tilde{p}_{\underline{Z}_{\mathcal{W}}^{(j)}|\underline{M}_{\mathcal{K}_s}^{(j)}}(\underline{z}^{(j)}|\underline{m}_{k_s}^{(j)})-\tilde{p}_{\underline{Z}_{\mathcal{W}}^{(j)}}(\underline{z}^{(j)})\right| \cdot \\
    & \quad\quad\quad\quad \prod_{q=1}^{j-1}\tilde{p}_{\underline{Z}_{\mathcal{W}}^{(q)}|\underline{M}_{\mathcal{K}_s}^{(q)}}(\underline{z}^{(q)}|\underline{m}_{k_s}^{(q)}) \cdot \prod_{k=j+1}^{\tilde{n}}\tilde{p}_{\underline{Z}_{\mathcal{W}}^{(k)}}(\underline{z}^{(k)}) \\
    & \quad \overset{(c)}{=} \sum_{j=1}^{\tilde{n}} \sum_{\underline{z}^{(j)}}\left|\tilde{p}_{\underline{Z}_{\mathcal{W}}^{(j)}|\underline{M}_{\mathcal{K}_s}^{(j)}}(\underline{z}^{(j)}|\underline{m}_{k_s}^{(j)})-\tilde{p}_{\underline{Z}_{\mathcal{W}}^{(j)}}(\underline{z}^{(j)})\right|\cdot \\
    & \quad\quad\quad \sum_{\underline{\hat{z}}^{(j)}}\prod_{q=1}^{j-1}\tilde{p}_{\underline{Z}_{\mathcal{W}}^{(q)}|\underline{M}_{\mathcal{K}_s}^{(q)}}(\underline{z}^{(q)}|\underline{m}_{k_s}^{(q)}) \cdot \prod_{k=i+1}^{\tilde{n}}\tilde{p}_{\underline{Z}_{\mathcal{W}}^{(k)}}(\underline{z}^{(k)}) \\
    & \quad = \sum_{j=1}^{\tilde{n}} \sum_{\underline{z}^{(j)}}\left|\tilde{p}_{\underline{Z}_{\mathcal{W}}^{(j)}|\underline{M}_{\mathcal{K}_s}^{(j)}}(\underline{z}^{(j)}|\underline{m}_{k_s}^{(j)})-\tilde{p}_{\underline{Z}_{\mathcal{W}}^{(j)}}(\underline{z}^{(j)})\right| \overset{(d)}{\leq} \tilde{n} \ell^{-\frac{t}{2}},
\end{aligned} 
\end{equation}
where (a) holds since $\tilde{p}_{\underline{Z}_{\mathcal{W}}|\underline{M}_{\mathcal{K}_s}}$ and $\tilde{p}_{\underline{Z}_{\mathcal{W}}}$ are induced from a completely uniform distribution, and (b) holds from \cite[Telescoping Expansion]{korada2010polar}. (c) is from denoting $\underline{\hat{z}}^{(j)}=\left(\underline{z}^{(1)},...,\underline{z}^{(j-1)},\underline{z}^{(j+1)},...,\underline{z}^{(\tilde{n})}\right)$. (d) holds according to \cite[Theorem 1]{SMSM}, by choosing $k_s \leq \ell - w - \ell\epsilon$, and $\ell\epsilon = \lceil t \log{\ell}\rceil$ for $t \geq 1$.

Now, we return to \eqref{FullSecrecy}. By substituting \eqref{eq:CondSec}-\eqref{eq:FullZ} with \eqref{eq:uniform-bound}-\eqref{eq:HelpLemmaBound}, we show that
\[
\mathbb{V}\left(p_{\underline{Z}_{\mathcal{W}} | \underline{M}_{\mathcal{K}_s}=\underline{m}_{k_s}},p_{\underline{Z}_{\mathcal{W}}}\right) \leq  \tilde{n} \ell^{-\frac{t}{2}} + 2 \sqrt{2 \ell \tilde{n} 2^{-n^\beta}}.
\]
\off{\begin{multline*}
\begin{aligned}
    &\mathbb{V}\left(p_{\underline{Z}_{\mathcal{W}} | \underline{M}_{K_s}=\underline{m}_{k_s}},p_{\underline{Z}_{\mathcal{W}}}\right) \\
    & \quad \leq \mathbb{V}\left(p_{\underline{Z}_{\mathcal{W}}|\underline{M}_{K_s}},\tilde{p}_{\underline{Z}_{\mathcal{W}}|\underline{M}_{K_s}}\right) +\\
    & \quad\quad\quad  \mathbb{V}\left(\tilde{p}_{\underline{Z}_{\mathcal{W}}|\underline{M}_{K_s}=\underline{m}_{k_s}},\tilde{p}_{\underline{Z}_{\mathcal{W}}}\right) + \mathbb{V}\left(\tilde{p}_{\underline{Z}_{\mathcal{W}}},p_{\underline{Z}_{\mathcal{W}}}\right) \\
    & \quad \leq \tilde{n} \cdot \ell^{-\frac{t}{2}} + 2\cdot \sqrt{2\cdot \ell \cdot \tilde{n} \cdot 2^{-n^\beta}}.
\end{aligned}
\end{multline*}}

We have shown that the channel scheme employed on $\underline{M}_{\mathcal{L}}$ is $k_s$-IS. To conclude the leakage proof against IT-Eve, we show that the scheme remains $k_s$-IS secured even when using the source coding scheme, by bounding $\mathbb{V}\left(p_{\underline{Z}_{\mathcal{W}} | \underline{V}_{\mathcal{K}_s}=\underline{v}_{k_s}},p_{\underline{Z}_{\mathcal{W}}}\right)$, and showing that for any $\underline{v}_{k_s} \in \underline{\mathcal{V}}_{\mathcal{K}_s}$ the
information leakage becomes negligible. Thus, we have
\begin{align*}
    &\mathbb{V}\left(p_{\underline{Z}_{\mathcal{W}} | \underline{V}_{\mathcal{K}_s}=\underline{v}_{k_s}},p_{\underline{Z}_{\mathcal{W}}}\right) \nonumber \\
    & \quad = \sum_{\underline{z}_w} \left|p_{\underline{Z}_{\mathcal{W}}|\underline{V}_{\mathcal{K}_s}}(\underline{z}_{w}|\underline{v}_{k_s})-p_{\underline{Z}_{\mathcal{W}}}(\underline{z}_{w})\right| \nonumber \\
    & \quad = \sum_{\underline{z}_{w}}\left|\sum_{\underline{m}_{k_s}} p_{\underline{Z}_{\mathcal{W}}|\underline{M}_{\mathcal{K}_s}\underline{V}_{\mathcal{K}_s}}(\underline{z}_{w}|\underline{m}_{k_s}\underline{v}_{k_s}) \cdot \right. \nonumber \\
    & \quad\quad\quad\quad\quad\quad\quad\quad\quad\quad \left. p_{\underline{M}_{\mathcal{K}_s}|\underline{V}_{\mathcal{K}_s}}(\underline{m}_{k_s}|\underline{v}_{k_s}) \right. \nonumber \\ 
    & \quad\quad\quad\quad\quad \left. - p_{\underline{Z}_{\mathcal{W}}}(\underline{z}_{w})\cdot \sum_{\underline{m}_{k_s}} p_{\underline{M}_{\mathcal{K}_s}|\underline{V}_{\mathcal{K}_s}}(\underline{m}_{k_s}|\underline{v}_{k_s})\right| \nonumber \\
    & \quad \overset{(a)}{\leq} \sum_{\underline{m}_{k_s}} \sum_{\underline{z}_{w}} p_{\underline{M}_{\mathcal{K}_s}|\underline{V}_{\mathcal{K}_s}}(\underline{m}_{k_s}|\underline{v}_{k_s}) \cdot \nonumber \\
    & \quad\quad\quad\quad\quad\quad \left|p_{\underline{Z}_{\mathcal{W}}|\underline{M}_{\mathcal{K}_s}\underline{V}_{\mathcal{K}_s}}(\underline{z}_{w}|\underline{m}_{k_s}\underline{v}_{k_s}) - p_{\underline{Z}_{\mathcal{W}}}(\underline{z}_{w})\right| \nonumber \\
    & \quad = \sum_{\underline{m}_{k_s}} p_{\underline{M}_{\mathcal{K}_s}|\underline{V}_{\mathcal{K}_s}}(\underline{m}_{k_s}|\underline{v}_{k_s}) \cdot \nonumber \\
    & \quad\quad\quad\quad\quad\quad \mathbb{V}\left(p_{\underline{Z}_{\mathcal{W}} | \underline{V}_{\mathcal{K}_s}=\underline{v}_{k_s}\underline{M}_{\mathcal{K}_s}=\underline{m}_{k_s}},p_{\underline{Z}_{\mathcal{W}}}\right) \nonumber \\
    & \quad \overset{(b)}{=} \sum_{\underline{m}_{\mathcal{K}_s}} p_{\underline{M}_{\mathcal{K}_s}|\underline{V}_{\mathcal{K}_s}}(\underline{m}_{k_s}|\underline{v}_{k_s}) \cdot \mathbb{V}\left(p_{\underline{Z}_{\mathcal{W}} | \underline{M}_{\mathcal{K}_s}=\underline{m}_{k_s}},p_{\underline{Z}_{\mathcal{W}}}\right) \\
    & \quad \leq \tilde{n} \ell^{-\frac{t}{2}} + 2 \sqrt{2 \ell \tilde{n}  2^{-n^\beta}},
\end{align*}
where (a) is from the triangle inequality, and (b) is since $\underline{V}_{\mathcal{K}_s} \rightarrow \underline{M}_{\mathcal{K}_s} \rightarrow \underline{Z}_{\mathcal{W}}$ is a Markov chain.
For the information leakage to be negligible, we give an upper and lower bound to $\ell$ compared to $n$. $\ell$ is lower bounded by $\omega(\tilde{n}^{\frac{2}{t}})$, s.t. the expression $\tilde{n} \ell^{-\frac{t}{2}}$ is negligible. In addition, $\ell$ is upper bounded by $o\left(2^{n^\beta}/\tilde{n}\right)$, s.t. the expression $2 \sqrt{2 \ell \tilde{n} 2^{-n^\beta}}$ is negligible. By choosing $\ell$ upholding both bounds, the information leakage to IT-Eve becomes negligible. This holds for any set of $\mathcal{K}_s \subset \mathcal{L}$  source messages, thus the code is $k_s$-IS.
\off{The lower bound of $\ell$ is required for $\tilde{n} \ell^{-\frac{t}{2}}$ to be negligible. This expression is negligible by taking $l$ to be any function of $n$ s.t. $\ell = \omega(\tilde{n}^{\frac{2}{t}})$. The upper bound on $l$ is required for $2\cdot \sqrt{2 \cdot \ell \cdot \tilde{n} \cdot 2^{-n^\beta}}$ to be negligible. This expression is negligible by taking $l$ to be any function of $\ell$ s.t. $\ell = o\left(2^{n^\beta}/\tilde{n}\right)$. By choosing $\ell$ s.t. those bounds hold, the information leakage to IT-Eve becomes negligible. 
This holds for any set of $\mathcal{K}_s \subset \mathcal{L}$  source messages, thus the code is $k_s$-IS.}

\section{Individual SS-CCA1 Against Crypto-Eve}\label{appendix:individual-ss-cca1}
We give here the full security proof of Theorem~\ref{Individual-SS-CCA1}. We aim to show each column $j \in \{1,...,\tilde{n}\}$ of the message matrix $\underline{M}_{\mathcal{L}}$ is ISS-CCA1. The proof is based on the equivalence between semantic security and indistinguishability \cite{goldwasser2019probabilistic}. We consider the maximal advantage for Crypto-Eve (maximal divination from uniform distribution probability), given the almost uniform messages after the source encoding stage (see Section~\ref{sec:main_results}) and prove that each column of the message matrix $\underline{M}_{\mathcal{L}}$ is IIND-CCA1 as given in \cite[Definition 4]{cohen2022partial}, and thus it is ISS-CCA1.

We assume the message matrix is the output of the source encoder from Appendix~\ref{appendix:src_code} with almost uniform messages. Alice and Crypto-Eve start playing the game as defined in \cite[definition 4]{cohen2022partial}. First, using the security parameter $c$, public and secret keys are created and denoted by $(p_c,s_c)$. We note that the security parameter $c$ is a function of the number of encrypted bits $1 \leq c < \ell$. Crypto-Eve sends a polynomial amount of ciphertexts to Alice and receives back their decryption. At this stage, Crypto-Eve chooses $i^{*} \in \{1,...,k_s\}$ (the case for $i^{*} \in \{k_s+1,...,l\}$ follows analogously), and two possible messages $M_{i^{*},1}$ and $M_{i^{*},2}$.

We add a step to the game that gives Crypto-Eve an additional advantage over Alice to show a stronger statement than in Definition~\ref{def:individuall-SS-CCA1}. Crypto-Eve is given the bits in positions $\{1,...,k_s\} \setminus i^{*}$. We show that still, Crypto-Eve is not able to distinguish between $M_{i^{*},1}$ and $M_{i^{*},2}$. Alice draws the bits in positions $\{k_s+1,...,\ell\}$ from the distribution induced by the source encoder, and chooses $h \in \{1,2\}$ uniformly at random s.t. the bit in position $i^{*}$ is $M_{i^{*},h}$. The message received is denoted by $M^\ell_h(j) \in \mathbb{F}_{2}^{\ell}$.

Alice encrypts $M^\ell_h(j)$ using $Crypt_2$ from \eqref{eq:encryption_crypt2}. First, Alice employs the IS channel code from Appendix~\ref{appendix:msg_encoder} to receive the encoded codeword denoted by $X^\ell_h(j) \in \mathbb{F}_{2}^{\ell}$. Second, Alice encrypts the first $c$ bits from $X^\ell_h(j)$ using $Crypt_1$. For the purpose of this proof, we denote the encrypted ciphertext by $\kappa = Crypt_2(M^\ell_h(j))$.

Upon receiving the ciphertext, $\kappa$, Crypto-Eve tries to guess $h$. First, $w = \ell - c$ of the bits from $\kappa$ are seen by Crypto-Eve as plaintext. Thus, Crypto-Eve can potentially reduce the number of possible codewords in each of the bins. \off{The average number of codewords that remain per bin after Crypto-Eve observes $w$ of the bits is $2^{\ell\epsilon}$ \cite{SMSM}, where $k_s = \ell - w - \ell\epsilon = c - \ell\epsilon$.} We denote the set of the remaining possible codewords from bin 1 and bin 2 by $\mathcal{B}_1$ and $\mathcal{B}_2$, respectively. It was shown in \cite{SMSM} that with high probability, the number of remaining possible codewords per bin deviates between $(1-\epsilon') 2^{\ell\epsilon} \leq |\mathcal{B}_1|,|\mathcal{B}_2| \leq (1+\epsilon') 2^{\ell\epsilon}$
for some $\epsilon' > 0$ and $\ell\epsilon$ s.t. $k_s = \ell - w -\ell\epsilon = c - \ell\epsilon$.

We assume the best possible scenario for Crypt-Eve s.t.: 1) The number of possible codewords remaining in bin 1 is as high as possible where the number of possible codewords remaining in bin 2 is as low as possible, $|\mathcal{B}_1| \geq |\mathcal{B}_2|$, 2) the induced probability of the codewords from bin 1 is as high as possible while the induced probability of the codewords from bin 2 is as low as possible \footnote{In \cite{cohen2022partial} the induced probabilities of each of the codewords were equal since the probability of the messages was uniform.}, and 3) the advantage Crypto-Eve has from $Crypt_1$ is as high as possible.

We start by bounding the maximum and minimum possible probabilities of the codewords in $\mathcal{B}_1$ and $\mathcal{B}_2$. We denote those probabilities by $p_{max}$ and $p_{min}$. First, we consider $j \in \mathcal{H}_V$, i.e. the bits in column $j$ were not padded by the uniform seed, and their entropy is lower bounded by $1 - \delta_n$. The case for $j \in \mathcal{J}_V$ will be addressed later on.

For each column $j$, we denote by $M_{k_w} \in \mathbb{F}_{2}^{k_w}$ the set of possible messages from bits $\{k_s+1,...,\ell\}$. In each column $j$ of the message matrix, the bits are independent since they are obtained from independent source messages. Thus, we have
\begin{align*}
    p(M_{k_w} =m_{k_w}) = \prod_{i=1}^{k_w}p(M_{k_w,i} = m_{k_w,i}),
\end{align*}
From \eqref{eq:unreliable-group}, we conclude that
\begin{equation*} \label{eq:bitEntropy}
    \begin{aligned}
        H(m_{k_w,i}) \geq H\left(m_{k_w,i}|m_{k_w,i}^{j-1}\right) \geq 1 - \delta_n,
    \end{aligned}
\end{equation*}
where $m_{k_w,i}^{j-1} = \left(m_{k_w,i}(1),...,m_{k_w,i}(j-1)\right)$ are the bits in columns $1$ to $j-1$ of the $i$-th row in the message matrix. We denote $\zeta \in (0,\frac{1}{2})$ s.t. $H(\frac{1}{2} - \zeta) = H(\frac{1}{2} + \zeta) = 1 - \delta_n$, and use $\zeta$ to bound $p_{min}$ and $p_{max}$
\begin{equation*}\label{eq:p-bounds}
    \begin{aligned}
        p_{min} \triangleq \left(\frac{1}{2} - \zeta\right)^{k_w} \leq p(\underline{m}_{k_w}) \leq \left(\frac{1}{2} + \zeta\right)^{k_w} \triangleq p_{max}.
    \end{aligned}
\end{equation*}
We bound $\zeta$ using the binary entropy bounds given in \cite{topsoe2001bounds}, $4p(1-p) \leq H_b(p) \leq \left(4p(1-p)\right)^{\ln{4}}$ for $0 < p < 1$. That is, by replacing $p = \frac{1}{2} + \zeta$ we obtain the upper bound by
\begin{equation*} \label{eq:right-side}
    \begin{aligned}
        H_b\left(\frac{1}{2} + \zeta\right) \leq \left(4\left(\frac{1}{2} + \zeta\right)\cdot \left(\frac{1}{2} - \zeta\right)\right)^{\frac{1}{\ln 4}},
    \end{aligned}
\end{equation*}
and the lower bound by
\begin{equation*} \label{eq:left-side}
    \begin{aligned}
        H_b\left(\frac{1}{2} + \zeta\right) \geq 4\left(\frac{1}{2} + \zeta\right)\left(\frac{1}{2} - \zeta\right).
    \end{aligned}
\end{equation*}
From which, we conclude
\begin{align*}
    \frac{1}{2} + \zeta \leq \frac{1}{2} + \frac{1}{2}\cdot \sqrt{1 - (1 - \delta_n)^{\ln 4}},
\end{align*}
and therefore, the upper bound for $\zeta$ is given by
\begin{equation}\label{eq:zeta-bound}
    \begin{aligned}
        \zeta \leq \frac{\sqrt{1 - (1 - \delta_n)^{\ln 4}}}{2}.
    \end{aligned}
\end{equation}

Now, we give the probability for some codeword from the set $\mathcal{B}_1 \cup \mathcal{B}_2$. We denote the codeword by $\alpha \in \mathcal{B}_1 \cup \mathcal{B}_2$. The probability for $\alpha$ is
\begin{equation} \label{eq:one-set}
    \begin{aligned}
        \mathbb{P}[\alpha | \alpha \in \mathcal{B}_1 \cup \mathcal{B}_2] = 
        \begin{cases}
            \frac{p_{max}}{|\mathcal{B}_1| \cdot p_{max} + |\mathcal{B}_2| \cdot p_{min}} & \text{if } \alpha \in \mathcal{B}_1 \\
            \frac{p_{min}}{|\mathcal{B}_1| \cdot p_{max} + |\mathcal{B}_2| \cdot p_{min}} & \text{if } \alpha \in \mathcal{B}_2
        \end{cases}
    \end{aligned}
\end{equation}
From \eqref{eq:one-set}, we conclude that the probability for a codeword $\alpha$ to be from the set $\mathcal{B}_1$ is
\begin{equation} \label{eq:bin1}
    \begin{aligned}
        \mathbb{P}[\alpha \in \mathcal{B}_1 | \alpha \in \mathcal{B}_1 \cup \mathcal{B}_2] = 
            \frac{|\mathcal{B}_1| \cdot p_{max}}{|\mathcal{B}_1| \cdot p_{max} + |\mathcal{B}_2| \cdot p_{min}}.
    \end{aligned}
\end{equation}

Since $Crypt_1$ is SS-CCA1, Crypto-Eve can obtain some negligible information about the original codeword or some function of the codeword. We consider the function $f: \{\mathcal{B}_1 \cup \mathcal{B}_2\} \rightarrow \{1,2\}$, i.e. the function that takes a codeword from the set $\{\mathcal{B}_1 \cup \mathcal{B}_2\}$ and outputs whether it belongs to $\mathcal{B}_1$ or $\mathcal{B}_2$
\begin{equation*} \label{eq:f(c*)}
    \begin{aligned}
        f(\alpha \in \{\mathcal{B}_1 \cup \mathcal{B}_2\}) = 
        \begin{cases}
            1 & \text{if }  \alpha \in \mathcal{B}_1\\
            2 & \text{if }  \alpha \in \mathcal{B}_2
        \end{cases}
    \end{aligned}
\end{equation*}
From the definition of SS-CCA1, we conclude
\begin{equation} \label{eq:P(c*)}
    \begin{aligned}
        \mathbb{P}[M_{i^{*},1}] =  \mathbb{P}[\alpha \in \mathcal{B}_1] + \epsilon_{ss-cca1},
    \end{aligned}
\end{equation}
where $\epsilon_{ss-cca1}$ is a negligible function in $c$ s.t. $\epsilon_{ss-cca1} \leq \frac{1}{c^{d}}$. Thus, by substituting \eqref{eq:bin1} into \eqref{eq:P(c*)}, we have
\begin{align}
    \mathbb{P}[M_{i^{*},1}] - \frac{1}{2} &=  \mathbb{P}[\alpha \in \mathcal{B}_1] + \epsilon_{ss-cca1} - \frac{1}{2} \nonumber \\
        \off{&= \frac{|\mathcal{B}|_1 \cdot p_{max}}{|\mathcal{B}_1| \cdot p_{max} + |\mathcal{B}_2| \cdot p_{min}} - \frac{1}{2} + \epsilon_{ss-cca1} \nonumber \\}
        &= \frac{|\mathcal{B}_1| \cdot p_{max} - |\mathcal{B}_2| \cdot p_{min}}{|\mathcal{B}_1| \cdot p_{max} + |\mathcal{B}_2| \cdot p_{min}} + \frac{1}{c^{d}} .\label{eq:adv-M}
\end{align}
We focus on bounding the left term of \eqref{eq:adv-M}
\begin{align}
    &\frac{|\mathcal{B}_1| \cdot p_{max} - |\mathcal{B}_2| \cdot p_{min}}{|\mathcal{B}_1| \cdot p_{max} + |\mathcal{B}_2| \cdot p_{min}} \nonumber \\
        &\overset{(a)}{=} \frac{(1+\epsilon') 2^{\ell\epsilon} p_{max} - (1-\epsilon') 2^{\ell\epsilon} p_{min}}{(1+\epsilon') 2^{\ell\epsilon} p_{max} + (1-\epsilon') 2^{\ell\epsilon} p_{min}} \nonumber \\
        &= \frac{1}{\frac{p_{max}+p_{min}}{p_{max}-p_{min}} + \epsilon'} + \frac{1}{\frac{p_{max}-p_{min}}{p_{max}+p_{min}} + \frac{1}{\epsilon'}}, \label{eq:p_max-p_min}
\end{align}
where (a) is from the assumption that $|\mathcal{B}_1| = (1+\epsilon')\cdot 2^{\ell\epsilon}$ and $|\mathcal{B}_2| = (1-\epsilon')\cdot 2^{\ell\epsilon}$. \off{By choosing $\epsilon' = \ell^{-t}$ s.t. $\ell\epsilon = \lceil t \log{\ell} \rceil$, and $t \geq 1$ \cite{SMSM},} We bound the term $\frac{p_{max}-p_{min}}{p_{max}+p_{min}}$ from \eqref{eq:p_max-p_min} by
\begin{align}
    \frac{p_{max}-p_{min}}{p_{max}+p_{min}} &= \frac{(\frac{1}{2}+\zeta)^{k_w}-(\frac{1}{2}-\zeta)^{k_w}}{(\frac{1}{2}+\zeta)^{k_w}+(\frac{1}{2}-\zeta)^{k_w}} \nonumber \\
    & = \frac{\sum_{i=0}^{k_w}\binom{k_w}{i}(2\zeta)^i - \sum_{i=0}^{k_w}\binom{k_w}{i}(-2\zeta)^i}{\sum_{i=0}^{k_w}\binom{k_w}{i}(2\zeta)^i + \sum_{i=0}^{k_w}\binom{k_w}{i}(-2\zeta)^i} \nonumber \\
    & = \frac{2\sum_{i=0}^{\lfloor \frac{k_w-1}{2} \rfloor}\binom{k_w}{2i+1}(2\zeta)^{2i+1}}{2\sum_{i=0}^{\lfloor \frac{k_w}{2} \rfloor}\binom{k_w}{2i}(2\zeta)^{2i}} \nonumber \\
    & \overset{(a)}{<} 2\zeta k_w \overset{(b)}{\leq} 2^{\frac{3}{2}}2^{\frac{-n^{\beta}}{2}}k_w, \label{eq:pmax-pmin-bound}
\end{align}
where (a) follows from the upper bound  $\frac{\sum_{i=0}^{\lfloor (k_w-1) / 2 \rfloor}\binom{k_w}{2i+1}(2\zeta)^{2i}}{\sum_{i=0}^{\lfloor k_w / 2 \rfloor}\binom{k_w}{2i}(2\zeta)^{2i}} < k_w$, which can be easily proved, and (b) is from \eqref{eq:zeta-bound}. By choosing $l \leq 2^{\frac{n^\beta}{2(t+1)}}$ s.t. $\ell\epsilon = \lceil t \log{\ell} \rceil$, for any $t \geq 1$, we have that $2^{\frac{3}{2}}2^{\frac{-n^{\beta}}{2}}k_w < 2^{\frac{3}{2}} \ell^{-t}$.

By substituting \eqref{eq:pmax-pmin-bound} into \eqref{eq:p_max-p_min} and taking $\epsilon' = \ell^{-t}$ \cite{SMSM}, it can be shown that for every $d'$ Crypto-Eve's advantage can be made smaller than $\frac{1}{c^{d'}}$ by choosing an appropriate $d$, large enough $\ell$ s.t $\ell \leq 2^{\frac{n^\beta}{2(t+1)}}$ and 
taking $t=\log{\ell}$. Thus, we showed that column $j \in \mathcal{H}_V$ is ISS-CCA1.

For the case of a column $j$ s.t. $j \in \mathcal{J}_V$, ISS-CCA1 follows directly from the SS-CCA1 of $Crypt_1$, considering Crypto-Eve has the maximal advantage in guessing the seed from its ciphertext. Thus, we can conclude the cryptosystem is IIND-CCA1 and thus ISS-CCA1 as well \cite{goldwasser2019probabilistic}.

\section{Seed Length}\label{appendix:seed_length}
The optimal size of the seed was defined in \cite{NegligbleCost,chou2013data}. For a message of size $n$ bits, the optimal size of the seed is 1) $d_{opt} = O(k_n\sqrt{n})$, $\forall k_n$ s.t. $\lim\limits_{n \rightarrow \infty} k_n \rightarrow \infty$, and 2) $d_{opt} = \Omega(\sqrt{n})$. However, it was shown by Chou et al. in \cite{NegligbleCost} that the seed size for the polar source coding scheme from Appendix~\ref{appendix:src_code} is sub-linear, i.e. $d_{opt} = o(n)$.\off{ and is considered sub-optimal.}
\off{However, the size of the seed for the polar codes source encoder/decoder from Appendix~\ref{appendix:src_code} is sub-optimal.}

\begin{figure}[htbp]
    \centering
    \includegraphics[width=0.95\linewidth]{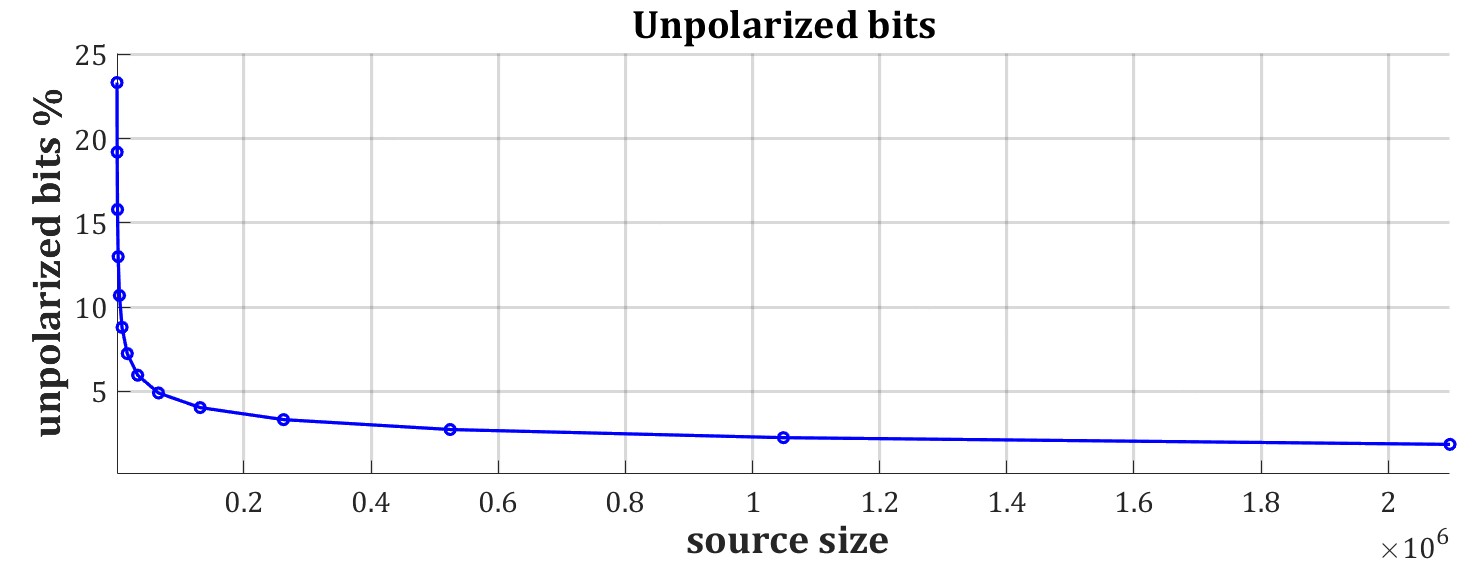}
    \caption{Numerical simulation of the seed size for a source $(\mathcal{V},p_V)$ with entropy $H(V) = 0.9$. For messages with a size greater than $2^{18}$ bits, the seed size already decreases to about $2.2\%$ of the compressed message size.}
    \label{fig:SeedSize}
\end{figure}

The length of the seed is directly affected by the polarization rate which determines the portion of the bits that remain unpolarized after the polarization transform, i.e. the size of the group $\mathcal{J}_v$. This size is often referred to as the gap to capacity \cite{hassani2010scaling,wang2023sub} since it determines how close the channel code gets to the optimal capacity of the communication channel. Polar codes for source and channel coding are directly linked, \cite{cronie2010lossless}, thus the gap to capacity is equal to the gap of the polar codes for source coding to the optimal compression rate \cite[Corollary 3.16]{wang2021complexity}. In \cite{hassani2010scaling} it was shown that the number of unpolarized bits can be bounded by
\vspace{-0.15cm}
\begin{equation*} \label{SeedBound}
    \begin{aligned}
        n^{0.7214} \leq d_{J} \leq n^{0.7331}.
    \end{aligned}
\end{equation*}
This bound is sub-optimal since for $k_n = n^{0.2}$, we have that $d_J = \Omega(n^{0.7})$. Yet, this overhead for the proposed NU-HUNCC scheme with non-uniform messages, for sufficiently large $n$, is still negligible compared to the compressed message size, as illustrated in Fig. \ref{fig:SeedSize}.
\vspace{-0.3cm}
\end{document}